\documentclass{paper}
\usepackage{graphicx}
\usepackage{graphicx,amssymb,amsmath,amsfonts,mathtext,color,wrapfig}

\def\be{\begin{eqnarray}}
\def\ee{\end{eqnarray}}

\hoffset= - 0.5in         

\begin{document}

\thispagestyle{empty}

\baselineskip14pt

\hfill ITEP/TH-45/11

\bigskip

\bigskip

\centerline{\Large{$\beta$-Deformation and Superpolynomials of $(n,m)$ Torus Knots
}}

\vspace{5ex}

\centerline{\large{\emph{Sh.Shakirov\footnote{shakirov@math.berkeley.edu}}}}

\vspace{5ex}

\centerline{\emph{Department of Mathematics, University of California, Berkeley, USA}}

\centerline{\emph{Center for Theoretical Physics, University of California, Berkeley, USA}}

\centerline{\emph{Institute for Theoretical and Experimental Physics, Moscow, Russia}}

\vspace{5ex}

\centerline{ABSTRACT}

\bigskip

{\footnotesize
Recent studies in several interrelated areas -- from combinatorics and representation theory in mathematics to quantum field theory and topological string theory in physics -- have independently revealed that many classical objects in these fields admit a relatively novel one-parameter deformation. This deformation, known in different contexts under the names of $\Omega$-background, refinement, or $\beta$-deformation, has a number of interesting mathematical implications. In particular, in Chern-Simons theory $\beta$-deformation transforms the classical HOMFLY invariants into Dunfield-Gukov-Rasmussen superpolynomials -- Poincare polynomials of a triply graded knot homology theory. As shown in arXiv:1106.4305, these superpolynomials are particular linear combinations of rational Macdonald dimensions,  distinguished by the polynomiality, integrality and positivity properties. We show that these properties alone do not fix the superpolynomials uniquely, by giving an example of a combination of Macdonald dimensions, that is always a positive integer polynomial but generally is not a superpolynomial.
}

\section{Introduction}

The mutual benefit of interaction between mathematics and physics is widely known and is proven by time. In some cases, as it happened with Einstein's theory of general relativity, a physical theory flourishes when a proper mathematical formulation is found. In other cases a breakthrough in a mathematical field occurs by application of physical ideas. Indeed, it is the interface with physics where some of the greatest developments in modern mathematics, like Witten's gauge theory construction of knot invariants \cite{WittenCS} or the discovery of mirror symmetry of Calabi-Yau manifolds \cite{Mirror1}-\cite{Mirror4}, were made.

Arguably one of the most interesting topics in modern mathematical physics is an emerging interplay between
numerous distinct fields, which were not initially thought to be connected. The central example of these connections is the AGT relation \cite{AGT} between instanton partition functions of 4d supersymmetric gauge theory \cite{Instanton1}-\cite{Instanton4} and conformal blocks of 2d conformal field theory \cite{CFT1}-\cite{CFT3}. The instanton partition functions contain as a limiting case \cite{Instanton3} the Seiberg-Witten prepotentials  \cite{SeibergWitten}, which are in correspondence with integrable systems \cite{SWInteg1}-\cite{NS}. Conformal blocks, in turn, are related to matrix models \cite{DFModels1}-\cite{DFModels3} which are naturally described in terms of symmetric functions and multivariate orthogonal polynomials -- the Schur, Jack and Macdonald functions \cite{Macdonald} -- and this paves a way into representation theory \cite{Hecke1}-\cite{CherednikDAHA} and combinatorics of Young diagrams \cite{Haiman1}-\cite{Gorsky2}. The last but not the least, all of this is embedded into string theory \cite{String1}-\cite{String3}.

The above short list is of course incomplete. It is intended only as a sketchy illustration of how different subjects provide complementary descriptions of one unified entity, like several coordinate charts provide different descriptions of a manifold in different regions. While it may be not completely clear yet what is this entity, and what is the underlying structure behind these connections, it is clearly worthwhile to develop and strengthen these links further, searching for a point of view from which the seemingly non-trivial identities become obvious.

From this perspective, it is especially important that each of the above-mentioned subjects contains a parameter, denoted as $\beta$, such that $\beta = 1$ is a distinguished point where things get simplified and a complete understanding can be reached. In conformal field theory that parameter $\beta$ determines the magnitude of the central charge $c$, via
$$
c = 1 - 6\left( \sqrt{\beta} - \dfrac{1}{\sqrt{\beta}} \right)^2
$$
so that the case $\beta = 1$ corresponds to conformal field theory with $c = 1$, which is known to be simple. In supersymmetric gauge theory, $\beta$ restricts the parameters $(\epsilon_1, \epsilon_2)$ of the $\Omega$-background used to regularize the integrals over instanton moduli spaces, via
$$
\beta = \dfrac{-\epsilon_1}{\epsilon_2}
$$
so that the case $\beta = 1$ corresponds to $\epsilon_1 + \epsilon_2 = 0$, which is again known to be the simplest case. In matrix models, $\beta$ determines the power of the Vandermonde determinant,
$$
\prod\limits_{i < j} (x_i - x_j)^{2\beta}
$$
which can be thought of as the relative strength of logarithmic Coulomb repulsion in the Coulomb gas picture \cite{DysonGas}, \cite{betamodel}. And again, $\beta = 1$ is known \cite{matrices1}, \cite{matrices2} to be the simplest case -- when the model actually reduces to integration over Hermitian matrices with eigenvalues $x_i$.

The same happens in all the other parts of the net of subjects described above: every time there exists a natural parameter $\beta$, and $\beta = 1$ corresponds to a distinguished, simplest, case. The relations between different subjects at this point also become simpler and can be often understood and proved by elementary means  \cite{beta1}, \cite{beta2}. The $\beta$-deformation, i.e. the deformation away from $\beta = 1$, is then the most interesting part of the story.

Recently, there has been an increase of interest to $\beta$-deformation of knot theory. Knots, their invariants and corresponding partition functions have been a part of mathematical physics ever since the work of Witten \cite{WittenCS} that expressed the HOMFLY knot invariants as Wilson loop averages in Chern-Simons theory. On the other hand, Chern-Simons theory is known to have relations to conformal field theory \cite{WittenCS} and from this perspective it is expectable \cite{AGT3d} that knot theory fits into the above general framework of AGT-like relations. Following this logic, one is inevitably led to conclusion that there should exist a certain distinguished one-parameter deformation of both Chern-Simons theory and of HOMFLY knot invariants.

Such a deformation of Chern-Simons theory was recently constructed in \cite{AS}. The results of \cite{AS} indicate that the corresponding deformation of HOMFLY invariants are the Dunfield-Gukov-Rasmussen \emph{knot superpolynomials} \cite{Superpoly} -- a one-parametric generalization of HOMFLY polynomials, which posess important connections to knot homology \cite{KnotHomology} and representation theory of doubly graded affine Hecke algebras \cite{Hecke2}-\cite{CherednikDAHA}. Because of this, the study of superpolynomials is interesting; unfortunately, their calculation is computationally not easy and hence not too many explicit formulas are available at the moment.

One general formula, based on refined Chern-Simons theory considerations, was given in \cite{AS} for (colored) superpolynomials of one particular family of knots -- the torus knots, that wind the surface of a torus $T^2$ with winding numbers $(n,m)$. The formula had a form
$$
\sum\limits_{Y, Y^{\prime}} \big( K_L \big)_{\varnothing, Y} {\cal N}_{R, Y^{\prime}}^{Y} \big( K_R \big)_{Y, \varnothing}
$$
where $R$ is the coloring representation, ${\cal N}_{R, Y^{\prime}}^{Y}$ are the refined Verlinde numbers and $K_L,K_R$ are certain modular matrices of refined Chern-Simons theory, see \cite{AS} for more details.

Despite very concrete and structured, this formula is not quite effective computationally, for large $n$ and $m$. Inspired by the ideas suggested in another recent paper \cite{ITEPKnots}, in this paper we search for another, more explicit and computationally effective formula for superpolynomials of all torus knots. As a guiding principle for this search, we take the polynomiality, integrality and positivity properties of the superpolynomials: these are obvious from the definition of superpolynomials as Poincare polynomials of some homology theory, but are non-trivial from the point of view of explicit construction \cite{ITEPKnots}. Though we do not succeed in finding such a formula, we give a natural generalization of a formula of \cite{ITEPKnots} that \emph{is} always a polynomial with positive integer coefficients. Surprisingly or not, this polynomial turns out to differ from the superpolynomial for sufficiently large $n, m$. This shows that by itself the polynomiality property is not too restrictive; more conditions should be added to fully describe the superpolynomials.

\section{$\beta$-deformation in knot theory}

\begin{wrapfigure}{l}{0.30\textwidth}
\vspace{-4ex}
  \begin{center}
    \includegraphics[width=0.23\textwidth]{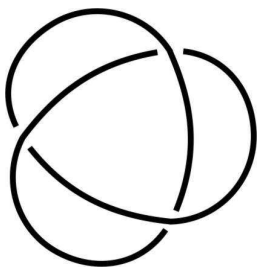}
  \end{center}
  \vspace{-2ex}
  \caption{A trefoil knot.}
  \vspace{-2ex}
\end{wrapfigure}
The knots that knot theory studies are embeddings $K: S^1 \rightarrow M$ of a circle into some manifold $M$, usually taken to be $M = S^3$ or simply $M = {\mathbb R}^3$. These quite simple geometric objects turned out to posess deep and non-trivial properties. A particular example of a knot, called the trefoil, is shown on Figure 1. Two knots are considered equivalent, if there exists a continious transformation in the embedding space that maps one into another. While this definition is very natural from topological point of view, it makes it very hard to check whether two knots are actually equivalent or not. This is known as \emph{recognition problem} in knot theory. It is generally hopeless to find the explicit transformation that connects two knots: for complicated enough knots one may spend a long time trying to transform one knot into another, or even to determine whether such a transformation exists.

Instead, it is natural to search for \emph{knot invariants}: certain quantities that can be computed for any given knot, and do not change if a knot is changed into an equivalent one. If two knots have different invariants, this immediately rules out the possibility that they are equivalent. However, if invariants are equal, this is not enough to conclude that the knots are equivalent. There could be many different knots with the same value of some invariant. The problem of the theory is thus to construct invariants that would distinguish any pair of different knots.

Starting from the beginning of knot theory, many simple knot invariants were constructed. Most widely known examples include the Alexander polynomial $\Delta(\textbf{q})$ \cite{Alexander} and the Jones polynomial $V(\textbf{q})$ \cite{Jones}. These invariants are polynomials in one auxillary variable $\textbf{q}$, and are quite rough, in a sence that many knots have the same Jones or Alexander polynomial. A more profound classical knot invariant, the HOMFLY polynomial $P(\textbf{a}, \textbf{q})$, was constructed later by a group of co-discoverers \cite{HOMFLY}. This invariant is already better in terms of its ability to distinguish knots. It depends on two parameters $\textbf{q}$ and $\textbf{a}$, in such a way that the Jones polynomial arises at $\textbf{a} = \textbf{q}^2$ and the Alexander polynomial at $\textbf{a} = \textbf{1}$.

It should be intuitively clear that, the more parameters the invariant contains, the better is it's ability to distinguish different knots -- since the amount of information, contained in the invariant, becomes larger. For this reason, construction of knot invariants that depend on more variables has always been an important task. On the basis of intuition coming from the AGT conjecture, a physicist would expect one such generalization -- the $\beta$-deformation -- to exist.

\begin{figure}
  \begin{center}
    \includegraphics[width=0.45\textwidth]{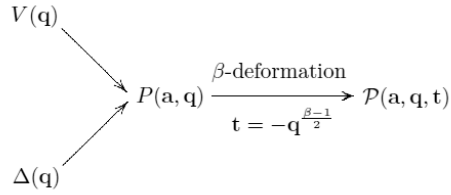}
  \end{center}
  \caption{The relations between knot invariants, and $\beta$-deformation of HOMFLY polynomials.}
\end{figure}
On the mathematical side, such a generalization was pointed out in the work of Dunfield, Gukov and Rasmussen. The new, triply-graded, invariants were called \emph{superpolynomials} and denoted ${\cal P}(\textbf{a},\textbf{q},\textbf{t})$. They depended on one additional variable $\textbf{t}$, and reduced to HOMFLY polynomials in particular case of $\textbf{t} = -1$. The relations between these four invariants are illustrated on Figure 2. For example, for the trefoil knot

$$
V(\textbf{q}) = 1 + \textbf{q}^4 - \textbf{q}^6 \ \ \ \ \ \Delta(\textbf{q}) = 1 - \textbf{q}^2 + \textbf{q}^4
$$
$$
\searrow \ \ \ \swarrow
$$
$$
P(\textbf{a}, \textbf{q}) = 1 + \textbf{q}^4 - \textbf{a}^2 \textbf{q}^2
$$
$$
\downarrow
$$
$$
{\cal P}(\textbf{a},\textbf{q},\textbf{t}) = 1 + \textbf{q}^4 \textbf{t}^2 + \textbf{a}^2 \textbf{q}^2 \textbf{t}^3
$$
\smallskip\\
Superpolynomials have a number of remarkable features; in particular, their coefficients are (as one can note on the above example) positive integers and, as such, they hint for existence of some combinatorial objects that they are counting. The objects are known as (triply graded) \emph{knot homologies} \cite{Superpoly}, \cite{KnotHomology}: along with their doubly graded Khovanov-Rozansky counterparts \cite{KhRoz}, they play the central role in modern mathematics of knots. The informal phrase "hint for existence of some combinatorial objects that they are counting" has been made precise by Khovanov in the framework of \emph{categorification} of combinatorial problems \cite{KnotHomology}.

As has been conjectured in \cite{AS}, superpolynomials are indeed the correct $\beta$-deformation of HOMFLY polynomials, i.e. the one that most naturally corresponds to $\beta$-deformed Chern-Simons theory, and hence fits appropriately into the AGT net of relations. However, superpolynomials have a drawback as well.
Namely, they are quite hard to compute. It is widely known that the HOMFLY polynomial can be defined and computed recursively, by going from bigger to simpler knots via the so-called skein relations. This gives a fast and reliable algorithm to find the HOMFLY polynomial. As of today, such a simple recursive procedure is not known for superpolynomials: instead, one needs to perform full-scale computation of knot homologies \cite{Superpoly}. As a rezult, practical calculation of these invariants is hard, even for the simplest knots. It goes without saying that in absence of an explicit formula, the understanding and, most importantly, practical applications of any theoretical concept remain limited.

\section{Torus knots: the case $\beta = 1$}

\begin{wrapfigure}{r}{0.30\textwidth}
\vspace{-4ex}
  \begin{center}
    \includegraphics[width=0.25\textwidth]{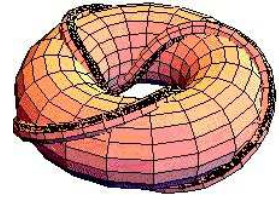}
  \end{center}
  \vspace{-4ex}
  \caption{A torus knot.}
  \vspace{-2ex}
\end{wrapfigure}
Instead of doing honest homological computations, one could follow another approach \cite{ITEPKnots}: start from the simple case of $\beta = 1$, where the answer is known explicitly, and try to figure out the proper deformation of that answer. In the paper \cite{ITEPKnots}, this task was formulated and partially solved for a particular family of knots: the torus knots. A torus knot is a knot that can be drawn on a surface of a torus without self-intersections; such knots are parametrized by two integer numbers, the winding numbers $(n \leq m)$. An example of a torus knot is given on Figure 3 (this is actually again the trefoil, with a pair of winding numbers $n = 2, m = 3$). For the knot to be one-component, $n$ and $m$ have to be relatively prime.

In the $\beta = 1$ case, the superpolynomial is nothing but the HOMFLY polynomial of a $(n,m)$ torus knot, which is known (see eq.(4.12) of \cite{TorusMM}) to be given explicitly by the formula
\begin{align}
P_{n,m}(\textbf{a},\textbf{q}) = {\rm const} \cdot \sum\limits_{|Y|=n} \big(T_Y\big)^{m/n} \ C_Y \ \chi_Y(1,q,q^2,\ldots,q^{N-1})
\label{OldFormula}
\end{align}
where the variables on the l.h.s. $\textbf{a},\textbf{q}$ and the variables on the r.h.s. $q, N$ are related via $\textbf{a} = q^N, \ \ \textbf{q} = q^2$ and ${\rm const}$ is an overall normalization, not important for us right now. The sum is taken over Young diagrams (partitions) $Y = (Y_1 \geq Y_2 \geq \ldots)$ of size $|Y| = Y_1 + Y_2 + \ldots = n$; the functions $\chi_Y(x_1, \ldots, x_N)$ are the Schur symmetric polynomials \cite{Macdonald} -- the characters of general linear groups, corresponding to irreducible representation $Y$; the numbers $C_Y$ are the coefficients of expansion of the $n$-th Newton power sum in Schur functions
\begin{align}
x_1^n + \ldots + x_N^n = \sum\limits_{|Y|=n} C_Y \ \chi_Y(x_1, \ldots, x_n)
\label{OldExpansion}
\end{align}
and $T_Y$ are the monomial quantities called framing factors:
\begin{align}
T_Y = \prod\limits_{(i,j) \in Y} q^{i-j} = q^{\sum_i (Y_i + {\widetilde Y}_i^2 - Y_i^2)/2}
\label{OldFraming}
\end{align}
The product here is taken over all cells $(i,j)$ of the Young diagram, viewed in the usual way as a collection of cells in a plane, and ${\widetilde Y}$ is the Young diagram transposed w.r.t the main diagonal. In other words, $(i,j)$ takes values in the set $\big\{ (i,j) \big| 1 \leq i \leq {\rm length}(Y), \ 1 \leq j \leq Y_i \big\}$.

\section{Torus knots: $\beta$-deformation}

Now that all the relevant ingredients for the $\beta = 1$ case are introduced, one can try to write down their proper $\beta$-deformations. The key observation on this path is that the formula (\ref{OldFormula}) is written in terms of Schur polynomials, which are the central algebraic objects of the classical representation theory of $GL(N)$ and associated theory of multivariate orthogonal polynomials. The natural $\beta$-deformation of Schur polynomials, as it is easy to guess, is either to Jack or to Macdonald polynomials -- but, as in present case the "quantum" parameter $q$ enters the formulas explicitly, one concludes that Macdonald polynomials are relevant:
\begin{align}
\chi_Y(x_1, \ldots, x_N) \ \ \ \mathop{\longrightarrow}^{\beta} \ \ \ M_Y(x_1, \ldots, x_N)
\end{align}
where the Macdonald parameters\footnote{For introduction to Schur, Jack and Macdonald polynomials, see \cite{MorozovUnitary} and the Appendices in \cite{beta1, beta2, BGW, AS}.} $t,q$ are taken such that $t = q^{\beta}$. Thus $\chi_Y(1,q,q^2,\ldots,q^{N-1})$, which is often called the \emph{quantum dimension}, gets substituted by the \emph{Macdonald dimension}
\begin{align}
\chi_Y(1,q,q^2,\ldots,q^{N-1}) \ \ \ \mathop{\longrightarrow}^{\beta} \ \ \ M_Y(1,t,t^2,\ldots,t^{N-1})
\end{align}
the Schur expansion coefficients $C_Y$ get substituted by the Macdonald expansion coefficients
\begin{align}
\sum_i x_i^n = \sum_{Y} C_Y \chi_Y \ \ \ \mathop{\longrightarrow}^{\beta} \ \ \ \sum_i x_i^n = \sum_{Y} C_Y M_Y
\end{align}
and the framing factors get substituted by the known Macdonald framing factors \cite{Framing1}-\cite{Framing3}, \cite{AS}:
\begin{align}
T_Y = \prod\limits_{(i,j) \in Y} q^{i-j} \ \ \ \mathop{\longrightarrow}^{\beta} \ \ \ T_Y = \prod\limits_{(i,j) \in Y} t^i / q^j
\end{align}
In this way one would arrive at the following naive conjecture
\begin{align}
\sum\limits_{|Y|=n} \big(T_Y\big)^{m/n} \ C_Y \ \chi_Y \ \mathop{\longrightarrow}^{?} \ \sum\limits_{|Y|=n} \big(T_Y\big)^{m/n} \ C_Y \ M_Y
\end{align}
However, as the investigation of \cite{ITEPKnots} has shown, this naive statement is false -- though the general logic is valid! In fact, what happens is that an additional quantity, called $\gamma_Y$ in \cite{ITEPKnots}, appears in the sum, that is hidden (trivial) in the $\beta = 1$ case:
\begin{align}
\sum\limits_{|Y|=n} \big(T_Y\big)^{m/n} \ C_Y \ \chi_Y \ \mathop{\longrightarrow}^{\beta} \ \sum\limits_{|Y|=n} \big(T_Y\big)^{m/n} \ C_Y \ \gamma^{(n,m)}_Y \ M_Y
\end{align}
The quantity $\gamma_Y$ is thus an essentially new ingredient, that is not seen at all at the point $\beta = 1$. For this reason, it is clear that evaluation of $\gamma_Y$ would not only be useful in the theory of superpolynomials; it could also shed some light on $\beta$-deformation in general. 

As noticed in \cite{ITEPKnots}, the factor $\gamma_Y$ can be often found from several restrictive properties that it should satisfy. Most importantly, the superpolynomial should be a polynomial, while each particular $M_Y(1,t,t^2,\ldots,t^{N-1})$ is of course a rational function. This means that $\gamma_Y$ have to be such, that all the poles of individual Macdonald dimensions disappear in the overall sum. In the case of $(n,nk+1)$ knots, this and some other conditions were enough to find $\gamma_Y$ explicitly,
\begin{align}
\gamma^{(n,nk+1)}_Y \sim \sum\limits_{(i,j) \in Y} t^i / q^j
\label{MorozovGamma}
\end{align}
while generalization to other $(n,m)$ remained an interesting problem. 

Here we present a generalization of (\ref{MorozovGamma}) that does satisfy the polynomiality property, as well as the integrality and positivity of coefficients. However, the rezult, despite it has these formal properties of superpolynomials and actually coincides with superpolynomials in simple enough cases (for all $(n,nk+1)$ and $(n,nk+n-1)$ knots), generally turns out to differ from the superpolynomial. This shows that, from Macdonald-theoretic point of view, superpolynomials are not distinguished by the above properties only.

Let $m = n k + r$ with $k$ the quotient and $r$ the remainder.
Let us define
\begin{align}
\boxed{ \ \
{\cal P}_{n,m}(\textbf{a},\textbf{q},\textbf{t}) = {\rm const} \cdot \sum\limits_{|Y|=n} \big(T_Y\big)^{k} \ C_Y \ \gamma_Y^{(r)} \ \dfrac{M_Y(1,t,t^2,\ldots,t^{N-1})}{M_{[1]}(1,t,t^2,\ldots,t^{N-1})}
 \ \ }
 \label{NewFormula}
\end{align}
where $q = \textbf{t}^2 \textbf{q}^2, \ t = \textbf{q}^2, \ t^N = -\textbf{a}^2 \textbf{t}$, the ${\rm const}$ is an overall normalization
\begin{align}
{\rm const} = \dfrac{1-q^{{\color{white} n}}}{1-q^n} \ t^{m} \ q^{rn + r(r-1)/2 - n(n-1)/2}
\label{NewNormalization}
\end{align}
quantities $C_Y$ are the expansion coefficients
\begin{align}
x_1^n + \ldots + x_N^n = \sum\limits_{|Y|=n} C_Y \ M_Y(x_1, \ldots, x_N)
\label{NewExpansion}
\end{align}
factors $T_Y$ are the Macdonald framing
\begin{align}
T_Y = \prod\limits_{(i,j) \in Y} t^{i}/q^{j} = t^{\sum_i (Y_i + {\widetilde Y}_i^2)/2} q^{-\sum_i (Y_i + Y_i^2)/2}
\label{NewFraming}
\end{align}
and factors $\gamma_Y$ are
\begin{align}
 \ \
\gamma_Y^{(r)} = e_{r}\left( \big\{ t^i / q^j\big\}_{(i,j) \in Y} \right)
 \ \
 \label{NewGamma}
\end{align}
where $e_r$ is the $r$-th elementary symmetric function of its arguments, i.e.

$$
\gamma_Y^{(1)} = \sum\limits_{(i,j) \in Y} t^i / q^j, \ \ \ \ \
\gamma_Y^{(2)} = \dfrac{1}{2!} \sum\limits_{(i,j) \neq (i^{\prime},j^{\prime}) \in Y} t^{i+i^{\prime}} / q^{j+j^{\prime}}
$$
$$
\gamma_Y^{(3)} = \dfrac{1}{3!} \sum\limits_{(i,j) \neq (i^{\prime},j^{\prime}) \neq (i^{\prime\prime},j^{\prime\prime}) \in Y} t^{i+i^{\prime}+i^{\prime\prime}} / q^{j+j^{\prime}+j^{\prime\prime}}
$$
\smallskip\\
and so on. This is the generalization of (\ref{MorozovGamma}) that we consider here.

\paragraph{Proposition 1.}${\cal P}_{n,m}$ is a polynomial with positive integer coefficients in $\textbf{a},\textbf{q},\textbf{t}$, if and only if $n$ and $m$ are relatively prime.

\paragraph{Proposition 2 (false).}If $n$ and $m$ are relatively prime, ${\cal P}_{n,m}(\textbf{a},\textbf{q},\textbf{t})$ is the superpolynomial of the $(n,m)$ torus knot. 

\paragraph{Proposition 3.} If $m = n k + 1$ or $m = nk + n-1$, then ${\cal P}_{n,m}(\textbf{a},\textbf{q},\textbf{t})$ is the superpolynomial of the corresponding torus knot. This was the original statement of \cite{ITEPKnots}.

\paragraph{}In the Examples section 8 below we list numerous examples of (\ref{NewFormula}) computed for various values of $n$ and $m$. All these examples support Proposition 1. Proposition 2, however, is false, and the first example in the list to show this is the (5,8) torus knot. It remains unclear whether the polynomials (\ref{NewFormula}) have interpretation as Poincare polynomials of any knot homology theory.

\pagebreak

\section{Explicitation of (\ref{NewFormula}): Macdonald dimension, and Cauchy identity}

To speed up calculations, it is desirable to explicitate various elements of (\ref{NewFormula}) as much, as possible. Macdonald theory provides enough opportunities to do so. For example, it is well-known \cite{Macdonald} that the Macdonald dimension is given by a closed-form product expression
\begin{align}
M_Y(1,t,t^2,\ldots,t^{N-1}) = t^{\sum_i ({\widetilde Y}_i^2 - Y_i)/2} \prod\limits_{(i,j) \in Y} \frac{ 1 - (t/q) t^{N-i} q^{j} }{1 - t^{{\widetilde Y}_j-i+1} q^{Y_i-j} }
\label{Dimension}
\end{align}
Note that, in the literature, the peculiar combinations $\mbox{Leg}_{i,j}(Y) = {\widetilde Y}_j-i$ and $\mbox{Arm}_{i,j}(Y) = Y_i-j$ are often called the leg-length and the arm-length. The other two relevant combinations, $\mbox{Coleg}_{i,j}(Y) = i-1$ and $\mbox{Coarm}_{i,j}(Y) = j-1$, are called the coleg-length and the coarm-length. We avoid this terminology in present paper. Putting $Y= [1]$, we find
\begin{align}
M_{[1]}(1,t,t^2,\ldots,t^{N-1}) = \dfrac{1 - t^{N}}{1 - t}
\label{DimensionSmall}
\end{align}
The above explicitation makes evaluation of $M_Y(1,t,t^2,\ldots,t^{N-1})$ a completely straightforward task. Evaluation of $\gamma_Y$ and $T_Y$ is by construction completely straightforward. What remains to explicitate is the coefficient $C_Y$ -- to avoid solving the equation (\ref{NewExpansion}). For this, let us use the Cauchy identity (a.k.a. completeness condition) for Macdonald polynomials \cite{Macdonald}:
\begin{align}
\sum\limits_{Y} m_Y \Lambda^{|Y|} M_Y(x_1, \ldots, x_N) M_Y(y_1, \ldots, y_L) = \exp\left( \sum\limits_{k = 1}^{\infty} \ \dfrac{\Lambda^k}{k} \ \dfrac{1 - t^k}{1 - q^k} \ p_k(x) p_k(y) \right)
\label{Cauchy}
\end{align}
where $p_k(x) =\sum_i x_i^k, p_k(y) = \sum_i y_i^k$ are the Newton power sums, and $m_Y$ is given by
\begin{align}
m_Y = \prod\limits_{(i,j) \in Y} \frac{1 - t^{Y^T_j-i+1} q^{Y_i-j} }{1 - t^{Y^T_j-i} q^{Y_i-j+1} }
\label{CauchyNorm}
\end{align}
Note, that the sum in the l.h.s. is taken over all Young diagrams, in the sence of formal power series in $\Lambda$. Note also, that the numbers of variables $N$ and $L$ are not related in any way and can be varied separately -- we are going to use that. If we now put
\begin{align}
(y_1, \ldots, y_L) = (1, t, t^2, \ldots, t^{L-1})
\end{align}
then
\begin{align}
p_k(y) = \dfrac{1 - t^{Lk}}{1 - t^k}
\end{align}
and we find
\begin{align}
\sum\limits_{Y} m_Y \Lambda^{|Y|} M_Y(x_1, \ldots, x_N) M_Y(1, t, t^2, \ldots, t^{L-1}) = \exp\left( \sum\limits_{k = 1}^{\infty} \ \dfrac{\Lambda^k}{k} \ \dfrac{1 - t^{Lk}}{1 - q^k} \ p_k(x) \right)
\end{align}
To obtain the desired expansion coefficients $C_Y$, we need to study a certain limit of this formula. Namely, consider the expression $\epsilon = 1 - t^L$. Since both sides of the Cauchy identity are formal \pagebreak power series, it makes sence to study the limit when $\epsilon \rightarrow 0$. In this limit, the exponent in the r.h.s. is equivalent to a linear term:

\begin{align}
\exp\left( \sum\limits_{k = 1}^{\infty} \ \dfrac{\Lambda^k}{k} \ \dfrac{1 - t^{Lk}}{1 - q^k} \ p_k(x) \right) = 1 + \sum\limits_{k = 1}^{\infty} \ \dfrac{\Lambda^k}{k} \ \dfrac{1 - t^{Lk}}{1 - q^k} \ p_k(x) + O(\epsilon^2)
\end{align}
\smallskip\\
Dividing both sides by $1 - t^L$ and taking the limit, we obtain

$$
\sum\limits_{Y} m_Y \Lambda^{|Y|} M_Y(x_1, \ldots, x_N) \left( \lim\limits_{t^L \rightarrow 1} \dfrac{M_Y(1, t, t^2, \ldots, t^{L-1})}{1 - t^L} \right) = \emph{}
$$
\begin{align}
\emph{} = \lim\limits_{t^L \rightarrow 1} \sum\limits_{k = 1}^{\infty} \ \dfrac{\Lambda^k}{k(1 - q^k)} \ \dfrac{1 - t^{Lk}}{1 - t^L} \ p_k(x)
\end{align}
\smallskip\\
In the r.h.s. the limit can be taken easily, and we find

\begin{align}
\sum\limits_{Y} m_Y \Lambda^{|Y|} M_Y(x_1, \ldots, x_N) \left( \lim\limits_{t^L \rightarrow 1} \dfrac{M_Y(1, t, t^2, \ldots, t^{L-1})}{1 - t^L} \right) = \sum\limits_{k = 1}^{\infty} \ \dfrac{\Lambda^k}{1 - q^k} \ p_k(x)
\end{align}
\smallskip\\
This allows us to conclude that

\begin{align}
p_k(x) = (1 - q^k) \sum\limits_{|Y|=k} m_Y \Lambda^{|Y|} M_Y(x_1, \ldots, x_N) \left( \lim\limits_{t^L \rightarrow 1} \dfrac{M_Y(1, t, t^2, \ldots, t^{L-1})}{1 - t^L} \right)
\end{align}
\smallskip\\
In this way we obtained the desired expression for the expansion coefficients $C_Y$:

\begin{align}
C_Y = (1-q^n) \ m_Y \ \left( \lim\limits_{t^L \rightarrow 1} \dfrac{M_Y(1, t, t^2, \ldots, t^{L-1})}{1 - t^L} \right)
\end{align}
\smallskip\\
Recalling that $M_Y(1, t, t^2, \ldots, t^{L-1})$ is given by the product (\ref{Dimension}), we easily compute that

\begin{align}
C_Y = (1-q^n) \ t^{\sum_i ({\widetilde Y}_i^2-Y_i)/2} \ \dfrac{ \prod\limits_{(i,j) \in Y \slash \{ [1,1]\}} 1 - (t/q) t^{-i} q^{j} }{ \prod\limits_{(i,j) \in Y} 1 - t^{Y^T_j-i} q^{Y_i-j+1} }
\label{ExpansionCoefficients}
\end{align}
\smallskip\\
which is the final expression for $C_Y$.

\pagebreak

\section{Examples}

In the examples below, we consider families of knots of the form $(n, nk + r)$ for fixed $n$ and $r$. For each family, we give a few examples of polynomials (\ref{NewFormula}) for first few $k$. The answers are polynomials with positive integer coefficients and coincide with superpolynomials known in the literature (e.g. with \cite{ITEPKnots}, \cite{Gorsky2}) in all cases except for (5,8), which is the first case not falling into $(n,nk+1)$ and $(n,nk+n-1)$ families.
  
It is also interesting to note that the dependence on $k$ is easy to describe by introducing a generating function
\begin{align}
{\cal F}_{n,r}\big(\textbf{a},\textbf{q},\textbf{t} \big| z \big) = \sum\limits_{k = 0}^{\infty} {\cal P}_{n, \ nk + r}(\textbf{a},\textbf{q},\textbf{t}) z^k
\end{align}
that describes in the most convenient way the full information about the family. For illustrative purposes, we include these generating functions in a few cases below.

\subsection*{The family (n,m) = (2,2k+1)}

\paragraph{The case $(n,m) = (2,3)$.} This is the trefoil. Eq. (\ref{NewFormula}) gives ${\cal P}_{2,3}(\textbf{a},\textbf{q},\textbf{t}) = 1+\textbf{q}^4 \textbf{t}^2 + \textbf{q}^2 \textbf{t}^3 \textbf{a}^2$. Since the rezults will soon become lengthy, from now on we switch to a more structured form of presenting the answers, by grouping the different terms w.r.t. their $\textbf{a}$-{\rm degree}:

\[
\begin{array}{c|lll}
\textbf{a}-{\rm degree} & {\rm coefficient} & \rule{0pt}{3mm}  \\
\hline \textbf{a}^0 & 1+\textbf{q}^4 \textbf{t}^2 & \rule{0pt}{5mm}  \\
\hline \textbf{a}^2 & \textbf{q}^2 \textbf{t}^3 & \rule{0pt}{5mm}  \\
\end{array}
\]

\paragraph{The case $(n,m) = (2,5)$}

\[
\begin{array}{c|lll}
\textbf{a}-{\rm degree} & {\rm coefficient} & \rule{0pt}{3mm}  \\
\hline \textbf{a}^0 & 1+\textbf{q}^4 \textbf{t}^2+\textbf{q}^8 \textbf{t}^4 & \rule{0pt}{5mm}  \\
\hline \textbf{a}^2 & \textbf{q}^2 \textbf{t}^3+\textbf{q}^6 \textbf{t}^5 & \rule{0pt}{5mm}  \\
\end{array}
\]

\paragraph{The case $(n,m) = (2,7)$}

\[
\begin{array}{c|lll}
\textbf{a}-{\rm degree} & {\rm coefficient} & \rule{0pt}{3mm}  \\
\hline \textbf{a}^0 & 1+\textbf{q}^4 \textbf{t}^2+\textbf{q}^8 \textbf{t}^4+\textbf{q}^{12} \textbf{t}^6 & \rule{0pt}{5mm}  \\
\hline \textbf{a}^2 & \textbf{q}^2 \textbf{t}^3+\textbf{q}^6 \textbf{t}^5+\textbf{q}^{10} \textbf{t}^7 & \rule{0pt}{5mm}  \\
\end{array}
\]

\paragraph{The $(2,2k+1)$ generating function}

\begin{align*}
{\cal F}_{2,1}\big(\textbf{a},\textbf{q},\textbf{t} \big| z \big) = \dfrac{1 + z \textbf{q}^2 \textbf{t}^3 \textbf{a}^2}{(1 - z)(1 - z \textbf{q}^4 \textbf{t}^2)}
\end{align*}

\subsection*{The family $(n,m) = (3,3k+1)$}

\paragraph{The case $(n,m) = (3,4)$}

\[
\begin{array}{c|lll}
\textbf{a}-{\rm degree} & {\rm coefficient} & \rule{0pt}{3mm}  \\
\hline \textbf{a}^0 & 1+\textbf{q}^4 \textbf{t}^2+\textbf{q}^6 \textbf{t}^4+\textbf{q}^8 \textbf{t}^4+\textbf{q}^{12} \textbf{t}^6 & \rule{0pt}{5mm}  \\
\hline \textbf{a}^2 & \textbf{q}^2 \textbf{t}^3+\textbf{q}^4 \textbf{t}^5+\textbf{q}^6 \textbf{t}^5+\textbf{q}^8 \textbf{t}^7+\textbf{q}^{10} \textbf{t}^7 & \rule{0pt}{5mm}  \\
\hline \textbf{a}^4 & \textbf{t}^8 \textbf{q}^6 & \rule{0pt}{5mm}  \\
\end{array}
\]

\paragraph{The case $(n,m) = (3,7)$}

\[
\begin{array}{c|lll}
\textbf{a}-{\rm degree} & {\rm coefficient} & \rule{0pt}{3mm}  \\
\hline \textbf{a}^0 & 1+\textbf{q}^4 \textbf{t}^2+\textbf{q}^6 \textbf{t}^4+\textbf{q}^8 \textbf{t}^4+\textbf{q}^{10} \textbf{t}^6+\textbf{q}^{12} \textbf{t}^6+\textbf{q}^{12} \textbf{t}^8+\textbf{q}^{14} \textbf{t}^8 +
& \rule{0pt}{5mm} \\ &
+ \textbf{q}^{16} \textbf{t}^8+\textbf{q}^{18} \textbf{t}^{10}+\textbf{q}^{20} \textbf{t}^{10}+\textbf{q}^{24} \textbf{t}^{12} & \rule{0pt}{5mm}  \\
\hline \textbf{a}^2 & \textbf{q}^2 \textbf{t}^3+\textbf{q}^4 \textbf{t}^5+\textbf{q}^6 \textbf{t}^5+2 \textbf{q}^8 \textbf{t}^7+\textbf{q}^{10} \textbf{t}^7+\textbf{q}^{10} \textbf{t}^9+2 \textbf{q}^{12} \textbf{t}^9 +
& \rule{0pt}{5mm} \\ &
+ \textbf{q}^{14} \textbf{t}^9+\textbf{q}^{14} \textbf{t}^{11}+2 \textbf{q}^{16} \textbf{t}^{11}+\textbf{q}^{18} \textbf{t}^{11}+\textbf{q}^{20} \textbf{t}^{13}+\textbf{q}^{22} \textbf{t}^{13} & \rule{0pt}{5mm}  \\
\hline \textbf{a}^4 & \textbf{q}^6 \textbf{t}^8+\textbf{q}^{10} \textbf{t}^{10}+\textbf{q}^{12} \textbf{t}^{12}+\textbf{q}^{14} \textbf{t}^{12}+\textbf{q}^{18} \textbf{t}^{14} & \rule{0pt}{5mm}  \\
\end{array}
\]

\paragraph{The case $(n,m) = (3,10)$}

\[
\begin{array}{c|lll}
\textbf{a}-{\rm degree} & {\rm coefficient} & \rule{0pt}{3mm}  \\
\hline \textbf{a}^0 & 1+\textbf{q}^4 \textbf{t}^2+\textbf{q}^6 \textbf{t}^4+\textbf{q}^8 \textbf{t}^4+\textbf{q}^{10} \textbf{t}^6+\textbf{q}^{12} \textbf{t}^6+\textbf{q}^{12} \textbf{t}^8+\textbf{q}^{14} \textbf{t}^8+
& \rule{0pt}{5mm} \\ &
+\textbf{q}^{16} \textbf{t}^8+\textbf{q}^{16} \textbf{t}^{10}+\textbf{q}^{18} \textbf{t}^{10}+\textbf{q}^{18} \textbf{t}^{12}+\textbf{q}^{20} \textbf{t}^{10}+\textbf{q}^{20} \textbf{t}^{12}+\textbf{q}^{22} \textbf{t}^{12}+
& \rule{0pt}{5mm} \\ &
+\textbf{q}^{24} \textbf{t}^{12}+\textbf{q}^{24} \textbf{t}^{14}+\textbf{q}^{26} \textbf{t}^{14}+\textbf{q}^{28} \textbf{t}^{14}+\textbf{q}^{30} \textbf{t}^{16}+\textbf{q}^{32} \textbf{t}^{16}+\textbf{q}^{36} \textbf{t}^{18} & \rule{0pt}{5mm}  \\
\hline \textbf{a}^2 & \textbf{q}^2 \textbf{t}^3+\textbf{q}^4 \textbf{t}^5+\textbf{q}^6 \textbf{t}^5+2 \textbf{q}^8 \textbf{t}^7+\textbf{q}^{10} \textbf{t}^7+\textbf{q}^{10} \textbf{t}^9+2 \textbf{q}^{12} \textbf{t}^9+\textbf{q}^{14} \textbf{t}^9+2 \textbf{q}^{14} \textbf{t}^{11}+
& \rule{0pt}{5mm} \\ &
+2 \textbf{q}^{16} \textbf{t}^{11}+\textbf{q}^{16} \textbf{t}^{13}+\textbf{q}^{18} \textbf{t}^{11}+2 \textbf{q}^{18} \textbf{t}^{13}+2 \textbf{q}^{20} \textbf{t}^{13}+\textbf{q}^{20} \textbf{t}^{15}+\textbf{q}^{22} \textbf{t}^{13}+
& \rule{0pt}{5mm} \\ &
+2 \textbf{q}^{22} \textbf{t}^{15}+2 \textbf{q}^{24} \textbf{t}^{15}+\textbf{q}^{26} \textbf{t}^{15}+\textbf{q}^{26} \textbf{t}^{17}+2 \textbf{q}^{28} \textbf{t}^{17}+\textbf{q}^{30} \textbf{t}^{17}+\textbf{q}^{32} \textbf{t}^{19}+\textbf{q}^{34} \textbf{t}^{19} & \rule{0pt}{5mm}  \\
\hline \textbf{a}^4 & \textbf{q}^6 \textbf{t}^8+\textbf{q}^{10} \textbf{t}^{10}+\textbf{q}^{12} \textbf{t}^{12}+\textbf{q}^{14} \textbf{t}^{12}+\textbf{q}^{16} \textbf{t}^{14}+\textbf{q}^{18} \textbf{t}^{14}+\textbf{q}^{18} \textbf{t}^{16}+
& \rule{0pt}{5mm} \\ &
+\textbf{q}^{20} \textbf{t}^{16}+\textbf{q}^{22} \textbf{t}^{16}+\textbf{q}^{24} \textbf{t}^{18}+\textbf{q}^{26} \textbf{t}^{18}+\textbf{q}^{30} \textbf{t}^{20} & \rule{0pt}{5mm}  \\
\end{array}
\]

\paragraph{The $(3,3k+1)$ generating function}

\begin{align*}
\nonumber {\cal F}_{3,1}\big(\textbf{a},\textbf{q},\textbf{t} \big| z \big) = \dfrac{1}{(1 - z)(1 - z \textbf{q}^6 \textbf{t}^4)(1 - z \textbf{q}^{12} \textbf{t}^{6})} \times \Big(1+(\textbf{q}^4 \textbf{t}^2+\textbf{q}^8 \textbf{t}^4) z+(\textbf{q}^2 \textbf{t}^3+\textbf{q}^4 \textbf{t}^5 + \\ + \textbf{q}^6 \textbf{t}^5+\textbf{q}^8 \textbf{t}^7+\textbf{q}^{10} \textbf{t}^7) z \textbf{a}^2+z^2 \textbf{a}^2 \textbf{q}^{12} \textbf{t}^9+z \textbf{a}^4 \textbf{q}^6 \textbf{t}^8+(\textbf{q}^{10} \textbf{t}^{10}+\textbf{q}^{14} \textbf{t}^{12}) z^2 \textbf{a}^4 \Big)
\end{align*}

\subsection*{The family $(n,m) = (3,3k+2)$}

\paragraph{The case $(n,m) = (3,5)$}

\[
\begin{array}{c|lll}
\textbf{a}-{\rm degree} & {\rm coefficient} & \rule{0pt}{3mm}  \\
\hline \textbf{a}^0 & 1+\textbf{q}^4 \textbf{t}^2+\textbf{q}^6 \textbf{t}^4+\textbf{q}^8 \textbf{t}^4+\textbf{q}^{10} \textbf{t}^6+\textbf{q}^{12} \textbf{t}^6+\textbf{q}^{16} \textbf{t}^8 & \rule{0pt}{5mm}  \\
\hline \textbf{a}^2 &\textbf{q}^2 \textbf{t}^3+\textbf{q}^4 \textbf{t}^5+\textbf{q}^6 \textbf{t}^5+2 \textbf{q}^8 \textbf{t}^7+\textbf{q}^{10} \textbf{t}^7+\textbf{q}^{12} \textbf{t}^9+\textbf{q}^{14} \textbf{t}^9 & \rule{0pt}{5mm}  \\
\hline \textbf{a}^4 & \textbf{q}^6 \textbf{t}^8+\textbf{q}^{10} \textbf{t}^{10} & \rule{0pt}{5mm}  \\
\end{array}
\]

\paragraph{The case $(n,m) = (3,8)$}

\[
\begin{array}{c|lll}
\textbf{a}-{\rm degree} & {\rm coefficient} & \rule{0pt}{3mm}  \\
\hline \textbf{a}^0 & 1+\textbf{q}^4 \textbf{t}^2+\textbf{q}^6 \textbf{t}^4+\textbf{q}^8 \textbf{t}^4+\textbf{q}^{10} \textbf{t}^6+\textbf{q}^{12} \textbf{t}^6+\textbf{q}^{12} \textbf{t}^8+\textbf{q}^{14} \textbf{t}^8 + & \rule{0pt}{5mm} \\ & + \textbf{q}^{16} \textbf{t}^8+\textbf{q}^{16} \textbf{t}^{10}+\textbf{q}^{18} \textbf{t}^{10}+\textbf{q}^{20} \textbf{t}^{10}+\textbf{q}^{22} \textbf{t}^{12}+\textbf{q}^{24} \textbf{t}^{12}+\textbf{q}^{28} \textbf{t}^{14} & \rule{0pt}{5mm}  \\
\hline \textbf{a}^2 & \textbf{q}^2 \textbf{t}^3+\textbf{q}^4 \textbf{t}^5+\textbf{q}^6 \textbf{t}^5+2 \textbf{q}^8 \textbf{t}^7+\textbf{q}^{10} \textbf{t}^7+\textbf{q}^{10} \textbf{t}^9+2 \textbf{q}^{12} \textbf{t}^9+\textbf{q}^{14} \textbf{t}^9+2 \textbf{q}^{14} \textbf{t}^{11}+ & \rule{0pt}{5mm} \\ & +2 \textbf{q}^{16} \textbf{t}^{11}+\textbf{q}^{18} \textbf{t}^{11}+\textbf{q}^{18} \textbf{t}^{13}+2 \textbf{q}^{20} \textbf{t}^{13}+\textbf{q}^{22} \textbf{t}^{13}+\textbf{q}^{24} \textbf{t}^{15}+\textbf{q}^{26} \textbf{t}^{15} & \rule{0pt}{5mm}  \\
\hline \textbf{a}^4 & \textbf{q}^6 \textbf{t}^8+\textbf{q}^{10} \textbf{t}^{10}+\textbf{q}^{12} \textbf{t}^{12}+\textbf{q}^{14} \textbf{t}^{12}+\textbf{q}^{16} \textbf{t}^{14}+\textbf{q}^{18} \textbf{t}^{14}+\textbf{q}^{22} \textbf{t}^{16} & \rule{0pt}{5mm}  \\
\end{array}
\]

\vspace{3ex}

\paragraph{The case $(n,m) = (3,11)$}

\[
\begin{array}{c|lll}
\textbf{a}-{\rm degree} & {\rm coefficient} & \rule{0pt}{3mm}  \\
\hline \textbf{a}^0 & 1+\textbf{q}^4 \textbf{t}^2+\textbf{q}^6 \textbf{t}^4+\textbf{q}^8 \textbf{t}^4+\textbf{q}^{10} \textbf{t}^6+\textbf{q}^{12} \textbf{t}^6+\textbf{q}^{12} \textbf{t}^8+\textbf{q}^{14} \textbf{t}^8+\textbf{q}^{16} \textbf{t}^8+\textbf{q}^{16} \textbf{t}^{10}+ & \rule{0pt}{5mm} \\ & +\textbf{q}^{18} \textbf{t}^{10}+\textbf{q}^{18} \textbf{t}^{12}+\textbf{q}^{20} \textbf{t}^{10}+\textbf{q}^{20} \textbf{t}^{12}+\textbf{q}^{22} \textbf{t}^{12}+\textbf{q}^{22} \textbf{t}^{14}+\textbf{q}^{24} \textbf{t}^{12}+\textbf{q}^{24} \textbf{t}^{14}+ & \rule{0pt}{5mm} \\ &+\textbf{q}^{26} \textbf{t}^{14} +\textbf{q}^{28} \textbf{t}^{14}+\textbf{q}^{28} \textbf{t}^{16}+\textbf{q}^{30} \textbf{t}^{16}+\textbf{q}^{32} \textbf{t}^{16}+\textbf{q}^{34} \textbf{t}^{18}+\textbf{q}^{36} \textbf{t}^{18}+\textbf{q}^{40} \textbf{t}^{20} & \rule{0pt}{5mm}  \\
\hline \textbf{a}^2 & \textbf{q}^2 \textbf{t}^3+\textbf{q}^4 \textbf{t}^5+\textbf{q}^6 \textbf{t}^5+2 \textbf{q}^8 \textbf{t}^7+\textbf{q}^{10} \textbf{t}^7+\textbf{q}^{10} \textbf{t}^9+2 \textbf{q}^{12} \textbf{t}^9+\textbf{q}^{14} \textbf{t}^9+2 \textbf{q}^{14} \textbf{t}^{11} + & \rule{0pt}{5mm} \\ & + 2 \textbf{q}^{16} \textbf{t}^{11}+\textbf{q}^{16} \textbf{t}^{13}+\textbf{q}^{18} \textbf{t}^{11}+2 \textbf{q}^{18} \textbf{t}^{13}+2 \textbf{q}^{20} \textbf{t}^{13}+2 \textbf{q}^{20} \textbf{t}^{15}+\textbf{q}^{22} \textbf{t}^{13}+ & \rule{0pt}{5mm} \\ & +2 \textbf{q}^{22} \textbf{t}^{15}+2 \textbf{q}^{24} \textbf{t}^{15}+\textbf{q}^{24} \textbf{t}^{17}+\textbf{q}^{26} \textbf{t}^{15}+2 \textbf{q}^{26} \textbf{t}^{17}+2 \textbf{q}^{28} \textbf{t}^{17}+\textbf{q}^{30} \textbf{t}^{17}+ & \rule{0pt}{5mm} \\ & +\textbf{q}^{30} \textbf{t}^{19}+2 \textbf{q}^{32} \textbf{t}^{19}+\textbf{q}^{34} \textbf{t}^{19}+\textbf{q}^{36} \textbf{t}^{21}+\textbf{q}^{38} \textbf{t}^{21} & \rule{0pt}{5mm}  \\
\hline \textbf{a}^4 & \textbf{q}^6 \textbf{t}^8+\textbf{q}^{10} \textbf{t}^{10}+\textbf{q}^{12} \textbf{t}^{12}+\textbf{q}^{14} \textbf{t}^{12}+\textbf{q}^{16} \textbf{t}^{14}+\textbf{q}^{18} \textbf{t}^{14}+\textbf{q}^{18} \textbf{t}^{16}+\textbf{q}^{20} \textbf{t}^{16}+ & \rule{0pt}{5mm} \\ & +\textbf{q}^{22} \textbf{t}^{16}+\textbf{q}^{22} \textbf{t}^{18}+\textbf{q}^{24} \textbf{t}^{18}+\textbf{q}^{26} \textbf{t}^{18}+\textbf{q}^{28} \textbf{t}^{20}+\textbf{q}^{30} \textbf{t}^{20}+\textbf{q}^{34} \textbf{t}^{22} & \rule{0pt}{5mm}  \\
\end{array}
\]

\vspace{3ex}

\paragraph{The $(3,3k+2)$ generating function}

\begin{align*}
\nonumber {\cal F}_{3,2}\big(\textbf{a},\textbf{q},\textbf{t} \big| z \big) = \dfrac{1}{(1 - z)(1 - z \textbf{q}^6 \textbf{t}^4)(1 - z \textbf{q}^{12} \textbf{t}^{6})} \times \Big(1+\textbf{q}^4 \textbf{t}^2+z \textbf{q}^8 \textbf{t}^4+\textbf{a}^2 \textbf{q}^2 \textbf{t}^3+\\ + (\textbf{q}^4 \textbf{t}^5 + \textbf{q}^6 \textbf{t}^5+ \textbf{q}^8 \textbf{t}^7 +\textbf{q}^{10} \textbf{t}^7+\textbf{q}^{12} \textbf{t}^9) z \textbf{a}^2+(\textbf{q}^6 \textbf{t}^8+\textbf{q}^{10} \textbf{t}^{10}) z \textbf{a}^4+\textbf{q}^{14} \textbf{t}^{12} z^2 \textbf{a}^4 \Big)
\end{align*}

\pagebreak

\subsection*{The family $(n,m) = (4,4k+1)$}

\paragraph{The case $(n,m) = (4,5)$}

\[
\begin{array}{c|lll}
\textbf{a}-{\rm degree} & {\rm coefficient} & \rule{0pt}{3mm}  \\
\hline \textbf{a}^0 & 1+\textbf{q}^4 \textbf{t}^2+\textbf{q}^6 \textbf{t}^4+\textbf{q}^8 \textbf{t}^4+\textbf{q}^8 \textbf{t}^6+\textbf{q}^{10} \textbf{t}^6+\textbf{q}^{12} \textbf{t}^6+\textbf{q}^{12} \textbf{t}^8+ & \rule{0pt}{5mm} \\ &+\textbf{q}^{14} \textbf{t}^8+\textbf{q}^{16} \textbf{t}^8+\textbf{q}^{16} \textbf{t}^{10}+\textbf{q}^{18} \textbf{t}^{10}+\textbf{q}^{20} \textbf{t}^{10}+\textbf{q}^{24} \textbf{t}^{12} & \rule{0pt}{5mm}  \\
\hline \textbf{a}^2 & \textbf{q}^2 \textbf{t}^3+\textbf{q}^4 \textbf{t}^5+\textbf{q}^6 \textbf{t}^5+\textbf{q}^6 \textbf{t}^7+2 \textbf{q}^8 \textbf{t}^7+\textbf{q}^{10} \textbf{t}^7+2 \textbf{q}^{10} \textbf{t}^9+2 \textbf{q}^{12} \textbf{t}^9+\textbf{q}^{12} \textbf{t}^{11}+ & \rule{0pt}{5mm} \\ &+\textbf{q}^{14} \textbf{t}^9+2 \textbf{q}^{14} \textbf{t}^{11}+2 \textbf{q}^{16} \textbf{t}^{11}+\textbf{q}^{18} \textbf{t}^{11}+\textbf{q}^{18} \textbf{t}^{13}+\textbf{q}^{20} \textbf{t}^{13}+\textbf{q}^{22} \textbf{t}^{13} & \rule{0pt}{5mm}  \\
\hline \textbf{a}^4 & \textbf{q}^6 \textbf{t}^8+\textbf{q}^8 \textbf{t}^{10}+\textbf{q}^{10} \textbf{t}^{10}+\textbf{q}^{10} \textbf{t}^{12}+\textbf{q}^{12} \textbf{t}^{12}+\textbf{q}^{14} \textbf{t}^{12}+\textbf{q}^{14} \textbf{t}^{14}+\textbf{q}^{16} \textbf{t}^{14}+\textbf{q}^{18} \textbf{t}^{14} & \rule{0pt}{5mm}  \\
\hline \textbf{a}^6 & \textbf{t}^{15} \textbf{q}^{12} & \rule{0pt}{5mm}  \\
\end{array}
\]

\paragraph{The case $(n,m) = (4,9)$}

\[
\begin{array}{c|lll}
\textbf{a}-{\rm degree} & {\rm coefficient} & \rule{0pt}{3mm}  \\
\hline \textbf{a}^0 & 1+\textbf{q}^4 \textbf{t}^2+\textbf{q}^6 \textbf{t}^4+\textbf{q}^8 \textbf{t}^4+\textbf{q}^8 \textbf{t}^6+\textbf{q}^{10} \textbf{t}^6+\textbf{q}^{12} \textbf{t}^6+2 \textbf{q}^{12} \textbf{t}^8+\textbf{q}^{14} \textbf{t}^8+\textbf{q}^{14} \textbf{t}^{10}+ & \rule{0pt}{5mm} \\ &+\textbf{q}^{16} \textbf{t}^8+2 \textbf{q}^{16} \textbf{t}^{10}+\textbf{q}^{16} \textbf{t}^{12}+\textbf{q}^{18} \textbf{t}^{10}+2 \textbf{q}^{18} \textbf{t}^{12}+\textbf{q}^{20} \textbf{t}^{10}+2 \textbf{q}^{20} \textbf{t}^{12}+\textbf{q}^{20} \textbf{t}^{14}+ & \rule{0pt}{5mm} \\ &+\textbf{q}^{22} \textbf{t}^{12}+2 \textbf{q}^{22} \textbf{t}^{14}+\textbf{q}^{24} \textbf{t}^{12}+2 \textbf{q}^{24} \textbf{t}^{14}+2 \textbf{q}^{24} \textbf{t}^{16}+\textbf{q}^{26} \textbf{t}^{14}+2 \textbf{q}^{26} \textbf{t}^{16}+\textbf{q}^{28} \textbf{t}^{14}+ & \rule{0pt}{5mm} \\ &+2 \textbf{q}^{28} \textbf{t}^{16}+\textbf{q}^{28} \textbf{t}^{18}+\textbf{q}^{30} \textbf{t}^{16}+2 \textbf{q}^{30} \textbf{t}^{18}+\textbf{q}^{32} \textbf{t}^{16}+2 \textbf{q}^{32} \textbf{t}^{18}+\textbf{q}^{32} \textbf{t}^{20}+\textbf{q}^{34} \textbf{t}^{18}+ & \rule{0pt}{5mm} \\ &+\textbf{q}^{34} \textbf{t}^{20}+\textbf{q}^{36} \textbf{t}^{18}+2 \textbf{q}^{36} \textbf{t}^{20}+\textbf{q}^{38} \textbf{t}^{20}+\textbf{q}^{40} \textbf{t}^{20}+\textbf{q}^{40} \textbf{t}^{22}+\textbf{q}^{42} \textbf{t}^{22}+\textbf{q}^{44} \textbf{t}^{22}+\textbf{q}^{48} \textbf{t}^{24} & \rule{0pt}{5mm}  \\
\hline \textbf{a}^2 & \textbf{q}^2 \textbf{t}^3+\textbf{q}^4 \textbf{t}^5+\textbf{q}^6 \textbf{t}^5+\textbf{q}^6 \textbf{t}^7+2 \textbf{q}^8 \textbf{t}^7+\textbf{q}^{10} \textbf{t}^7+3 \textbf{q}^{10} \textbf{t}^9+2 \textbf{q}^{12} \textbf{t}^9+2 \textbf{q}^{12} \textbf{t}^{11}+\textbf{q}^{14} \textbf{t}^9+ & \rule{0pt}{5mm} \\ &+4 \textbf{q}^{14} \textbf{t}^{11}+\textbf{q}^{14} \textbf{t}^{13}+2 \textbf{q}^{16} \textbf{t}^{11}+4 \textbf{q}^{16} \textbf{t}^{13}+\textbf{q}^{18} \textbf{t}^{11}+4 \textbf{q}^{18} \textbf{t}^{13}+3 \textbf{q}^{18} \textbf{t}^{15}+2 \textbf{q}^{20} \textbf{t}^{13}+ & \rule{0pt}{5mm} \\ &+5 \textbf{q}^{20} \textbf{t}^{15}+\textbf{q}^{22} \textbf{t}^{13}+\textbf{q}^{20} \textbf{t}^{17}+4 \textbf{q}^{22} \textbf{t}^{15}+4 \textbf{q}^{22} \textbf{t}^{17}+2 \textbf{q}^{24} \textbf{t}^{15}+5 \textbf{q}^{24} \textbf{t}^{17}+\textbf{q}^{26} \textbf{t}^{15}+ & \rule{0pt}{5mm} \\ &+\textbf{q}^{24} \textbf{t}^{19}+4 \textbf{q}^{26} \textbf{t}^{17}+4 \textbf{q}^{26} \textbf{t}^{19}+2 \textbf{q}^{28} \textbf{t}^{17}+5 \textbf{q}^{28} \textbf{t}^{19}+\textbf{q}^{30} \textbf{t}^{17}+\textbf{q}^{28} \textbf{t}^{21}+4 \textbf{q}^{30} \textbf{t}^{19}+ & \rule{0pt}{5mm} \\ &+3 \textbf{q}^{30} \textbf{t}^{21}+2 \textbf{q}^{32} \textbf{t}^{19}+4 \textbf{q}^{32} \textbf{t}^{21}+\textbf{q}^{34} \textbf{t}^{19}+4 \textbf{q}^{34} \textbf{t}^{21}+\textbf{q}^{34} \textbf{t}^{23}+2 \textbf{q}^{36} \textbf{t}^{21}+2 \textbf{q}^{36} \textbf{t}^{23}+ & \rule{0pt}{5mm} \\ &+\textbf{q}^{38} \textbf{t}^{21}+3 \textbf{q}^{38} \textbf{t}^{23}+2 \textbf{q}^{40} \textbf{t}^{23}+\textbf{q}^{42} \textbf{t}^{23}+\textbf{q}^{42} \textbf{t}^{25}+\textbf{q}^{44} \textbf{t}^{25}+\textbf{q}^{46} \textbf{t}^{25} & \rule{0pt}{5mm}  \\
\hline \textbf{a}^4 & \textbf{q}^6 \textbf{t}^8+\textbf{q}^8 \textbf{t}^{10}+\textbf{q}^{10} \textbf{t}^{10}+\textbf{q}^{10} \textbf{t}^{12}+2 \textbf{q}^{12} \textbf{t}^{12}+\textbf{q}^{14} \textbf{t}^{12}+3 \textbf{q}^{14} \textbf{t}^{14}+2 \textbf{q}^{16} \textbf{t}^{14}+2 \textbf{q}^{16} \textbf{t}^{16}+ & \rule{0pt}{5mm} \\ &+\textbf{q}^{18} \textbf{t}^{14}+4 \textbf{q}^{18} \textbf{t}^{16}+\textbf{q}^{18} \textbf{t}^{18}+2 \textbf{q}^{20} \textbf{t}^{16}+3 \textbf{q}^{20} \textbf{t}^{18}+\textbf{q}^{22} \textbf{t}^{16}+4 \textbf{q}^{22} \textbf{t}^{18}+2 \textbf{q}^{22} \textbf{t}^{20}+ & \rule{0pt}{5mm} \\ &+2 \textbf{q}^{24} \textbf{t}^{18}+4 \textbf{q}^{24} \textbf{t}^{20}+\textbf{q}^{26} \textbf{t}^{18}+4 \textbf{q}^{26} \textbf{t}^{20}+2 \textbf{q}^{26} \textbf{t}^{22}+2 \textbf{q}^{28} \textbf{t}^{20}+3 \textbf{q}^{28} \textbf{t}^{22}+\textbf{q}^{30} \textbf{t}^{20}+ & \rule{0pt}{5mm} \\ &+4 \textbf{q}^{30} \textbf{t}^{22}+\textbf{q}^{30} \textbf{t}^{24}+2 \textbf{q}^{32} \textbf{t}^{22}+2 \textbf{q}^{32} \textbf{t}^{24}+\textbf{q}^{34} \textbf{t}^{22}+3 \textbf{q}^{34} \textbf{t}^{24}+2 \textbf{q}^{36} \textbf{t}^{24}+\textbf{q}^{38} \textbf{t}^{24}+ & \rule{0pt}{5mm} \\ &+\textbf{q}^{38} \textbf{t}^{26}+\textbf{q}^{40} \textbf{t}^{26}+\textbf{q}^{42} \textbf{t}^{26} & \rule{0pt}{5mm}  \\
\hline \textbf{a}^6 & \textbf{q}^{12} \textbf{t}^{15}+\textbf{q}^{16} \textbf{t}^{17}+\textbf{q}^{18} \textbf{t}^{19}+\textbf{q}^{20} \textbf{t}^{19}+\textbf{q}^{20} \textbf{t}^{21}+\textbf{q}^{22} \textbf{t}^{21}+\textbf{q}^{24} \textbf{t}^{21}+\textbf{q}^{24} \textbf{t}^{23}+\textbf{q}^{26} \textbf{t}^{23}+ & \rule{0pt}{5mm} \\ &+\textbf{q}^{28} \textbf{t}^{23}+\textbf{q}^{28} \textbf{t}^{25}+\textbf{q}^{30} \textbf{t}^{25}+\textbf{q}^{32} \textbf{t}^{25}+\textbf{q}^{36} \textbf{t}^{27} & \rule{0pt}{5mm}  \\
\end{array}
\]

\pagebreak

\paragraph{The $(4,4k+1)$ generating function}

\begin{align*}
{\cal F}_{4,1}\big(\textbf{a},\textbf{q},\textbf{t} \big| z \big) = \dfrac{1}{(1 - z)(1 - z \textbf{q}^8 \textbf{t}^6)(1 - z \textbf{q}^{12} \textbf{t}^{8})(1 - z \textbf{q}^{16} \textbf{t}^{10})(1 - z \textbf{q}^{24} \textbf{t}^{12})} \times \end{align*}
\begin{center}
$\times \Big(1+(\textbf{q}^4 \textbf{t}^2+\textbf{q}^6 \textbf{t}^4+\textbf{q}^8 \textbf{t}^4+\textbf{q}^{10} \textbf{t}^6+\textbf{q}^{12} \textbf{t}^6+\textbf{q}^{14} \textbf{t}^8+\textbf{q}^{16} \textbf{t}^8+\textbf{q}^{18} \textbf{t}^{10}+\textbf{q}^{20} \textbf{t}^{10}) z+(-\textbf{q}^{16} \textbf{t}^{10}-\textbf{q}^{20} \textbf{t}^{12}+\textbf{q}^{22} \textbf{t}^{12}-\textbf{q}^{22} \textbf{t}^{14}-\textbf{q}^{24} \textbf{t}^{14}+\textbf{q}^{26} \textbf{t}^{14}-\textbf{q}^{26} \textbf{t}^{16}-\textbf{q}^{28} \textbf{t}^{16}-\textbf{q}^{32} \textbf{t}^{18}) z^2+(\textbf{q}^2 \textbf{t}^3+\textbf{q}^4 \textbf{t}^5+\textbf{q}^6 \textbf{t}^5+\textbf{q}^6 \textbf{t}^7+2 \textbf{q}^8 \textbf{t}^7+\textbf{q}^{10} \textbf{t}^7+2 \textbf{q}^{10} \textbf{t}^9+2 \textbf{q}^{12} \textbf{t}^9+\textbf{q}^{12} \textbf{t}^{11}+\textbf{q}^{14} \textbf{t}^9+2 \textbf{q}^{14} \textbf{t}^{11}+2 \textbf{q}^{16} \textbf{t}^{11}+\textbf{q}^{18} \textbf{t}^{11}+\textbf{q}^{18} \textbf{t}^{13}+\textbf{q}^{20} \textbf{t}^{13}+\textbf{q}^{22} \textbf{t}^{13}) z \textbf{a}^2+(-\textbf{q}^{30} \textbf{t}^{18}-\textbf{q}^{34} \textbf{t}^{20}-\textbf{q}^{36} \textbf{t}^{20}-\textbf{q}^{38} \textbf{t}^{22}-\textbf{q}^{42} \textbf{t}^{24}) z^3+(\textbf{q}^{16} \textbf{t}^{13}+\textbf{q}^{20} \textbf{t}^{13}+\textbf{q}^{22} \textbf{t}^{15}-\textbf{q}^{22} \textbf{t}^{17}+2 \textbf{q}^{24} \textbf{t}^{15}-\textbf{q}^{24} \textbf{t}^{17}+\textbf{q}^{26} \textbf{t}^{17}-\textbf{q}^{26} \textbf{t}^{19}+\textbf{q}^{28} \textbf{t}^{17}+\textbf{q}^{32} \textbf{t}^{21}) z^2 \textbf{a}^2+(\textbf{q}^6 \textbf{t}^8+\textbf{q}^8 \textbf{t}^{10}+\textbf{q}^{10} \textbf{t}^{10}+\textbf{q}^{10} \textbf{t}^{12}+\textbf{q}^{12} \textbf{t}^{12}+\textbf{q}^{14} \textbf{t}^{12}+\textbf{q}^{14} \textbf{t}^{14}+\textbf{q}^{16} \textbf{t}^{14}+\textbf{q}^{18} \textbf{t}^{14}) z \textbf{a}^4+(-\textbf{q}^{26} \textbf{t}^{19}-\textbf{q}^{28} \textbf{t}^{19}-\textbf{q}^{28} \textbf{t}^{21}-2 \textbf{q}^{30} \textbf{t}^{21}-2 \textbf{q}^{32} \textbf{t}^{21}-\textbf{q}^{32} \textbf{t}^{23}-\textbf{q}^{34} \textbf{t}^{21}-2 \textbf{q}^{34} \textbf{t}^{23}-2 \textbf{q}^{36} \textbf{t}^{23}-\textbf{q}^{36} \textbf{t}^{25}-\textbf{q}^{38} \textbf{t}^{23}-2 \textbf{q}^{38} \textbf{t}^{25}-2 \textbf{q}^{40} \textbf{t}^{25}-\textbf{q}^{40} \textbf{t}^{27}-2 \textbf{q}^{42} \textbf{t}^{27}-\textbf{q}^{44} \textbf{t}^{27}-\textbf{q}^{44} \textbf{t}^{29}-\textbf{q}^{46} \textbf{t}^{29}) z^3 \textbf{a}^2+(\textbf{q}^{12} \textbf{t}^{12}+\textbf{q}^{14} \textbf{t}^{14}+\textbf{q}^{16} \textbf{t}^{14}+\textbf{q}^{16} \textbf{t}^{16}+2 \textbf{q}^{18} \textbf{t}^{16}+2 \textbf{q}^{20} \textbf{t}^{16}+\textbf{q}^{20} \textbf{t}^{18}+\textbf{q}^{22} \textbf{t}^{16}+\textbf{q}^{22} \textbf{t}^{18}+2 \textbf{q}^{24} \textbf{t}^{18}+\textbf{q}^{24} \textbf{t}^{20}+\textbf{q}^{26} \textbf{t}^{18}+\textbf{q}^{26} \textbf{t}^{20}+2 \textbf{q}^{28} \textbf{t}^{20}+\textbf{q}^{28} \textbf{t}^{22}+2 \textbf{q}^{30} \textbf{t}^{22}+\textbf{q}^{32} \textbf{t}^{22}+\textbf{q}^{32} \textbf{t}^{24}+\textbf{q}^{34} \textbf{t}^{24}+\textbf{q}^{36} \textbf{t}^{24}) z^2 \textbf{a}^4-z^4 \textbf{a}^2 \textbf{q}^{48} \textbf{t}^{31}+z \textbf{a}^6 \textbf{q}^{12} \textbf{t}^{15}+(-\textbf{q}^{24} \textbf{t}^{20}-\textbf{q}^{26} \textbf{t}^{22}-2 \textbf{q}^{28} \textbf{t}^{22}-\textbf{q}^{30} \textbf{t}^{22}-2 \textbf{q}^{30} \textbf{t}^{24}-2 \textbf{q}^{32} \textbf{t}^{24}-\textbf{q}^{32} \textbf{t}^{26}-\textbf{q}^{34} \textbf{t}^{24}-2 \textbf{q}^{34} \textbf{t}^{26}-\textbf{q}^{36} \textbf{t}^{24}-\textbf{q}^{36} \textbf{t}^{26}-\textbf{q}^{36} \textbf{t}^{28}-\textbf{q}^{38} \textbf{t}^{26}-2 \textbf{q}^{38} \textbf{t}^{28}-2 \textbf{q}^{40} \textbf{t}^{28}-\textbf{q}^{40} \textbf{t}^{30}-\textbf{q}^{42} \textbf{t}^{28}-2 \textbf{q}^{42} \textbf{t}^{30}-2 \textbf{q}^{44} \textbf{t}^{30}-\textbf{q}^{46} \textbf{t}^{32}-\textbf{q}^{48} \textbf{t}^{32}) z^3 \textbf{a}^4+(\textbf{q}^{16} \textbf{t}^{17}+\textbf{q}^{18} \textbf{t}^{19}+\textbf{q}^{20} \textbf{t}^{19}+\textbf{q}^{22} \textbf{t}^{21}+\textbf{q}^{24} \textbf{t}^{21}+\textbf{q}^{26} \textbf{t}^{23}+\textbf{q}^{28} \textbf{t}^{23}+\textbf{q}^{30} \textbf{t}^{25}+\textbf{q}^{32} \textbf{t}^{25}) z^2 \textbf{a}^6+(-\textbf{q}^{44} \textbf{t}^{32}-\textbf{q}^{46} \textbf{t}^{32}-\textbf{q}^{48} \textbf{t}^{34}-\textbf{q}^{50} \textbf{t}^{34}-\textbf{q}^{52} \textbf{t}^{36}) z^4 \textbf{a}^4+(-\textbf{q}^{28} \textbf{t}^{25}-\textbf{q}^{32} \textbf{t}^{27}+\textbf{q}^{34} \textbf{t}^{27}-\textbf{q}^{34} \textbf{t}^{29}-\textbf{q}^{36} \textbf{t}^{29}+\textbf{q}^{38} \textbf{t}^{29}-\textbf{q}^{38} \textbf{t}^{31}-\textbf{q}^{40} \textbf{t}^{31}-\textbf{q}^{44} \textbf{t}^{33}) z^3 \textbf{a}^6+(-\textbf{q}^{42} \textbf{t}^{33}-\textbf{q}^{46} \textbf{t}^{35}-\textbf{q}^{48} \textbf{t}^{35}-\textbf{q}^{50} \textbf{t}^{37}-\textbf{q}^{54} \textbf{t}^{39}) z^4 \textbf{a}^6\Big)$
\end{center}
As one can see, explicit expressions for generating functions become increasingly lengthy when $n$ increases; we omit them from now on. This does not necessarily mean that such generating functions do not posess nice properties: a better representation for them might exist.

\subsection*{The family $(n,m) = (4,4k+3)$}

\paragraph{The case $(n,m) = (4,7)$}

\[
\begin{array}{c|lll}
\textbf{a}-{\rm degree} & {\rm coefficient} & \rule{0pt}{3mm}  \\
\hline \textbf{a}^0 & 1+\textbf{q}^4 \textbf{t}^2+\textbf{q}^6 \textbf{t}^4+\textbf{q}^8 \textbf{t}^4+\textbf{q}^8 \textbf{t}^6+\textbf{q}^{10} \textbf{t}^6+\textbf{q}^{12} \textbf{t}^6+2 \textbf{q}^{12} \textbf{t}^8+\textbf{q}^{14} \textbf{t}^8+\textbf{q}^{14} \textbf{t}^{10}+\textbf{q}^{16} \textbf{t}^8+ & \rule{0pt}{5mm} \\ &+2 \textbf{q}^{16} \textbf{t}^{10}+\textbf{q}^{18} \textbf{t}^{10}+\textbf{q}^{18} \textbf{t}^{12}+\textbf{q}^{20} \textbf{t}^{10}+2 \textbf{q}^{20} \textbf{t}^{12}+\textbf{q}^{22} \textbf{t}^{12}+\textbf{q}^{22} \textbf{t}^{14}+\textbf{q}^{24} \textbf{t}^{12}+ & \rule{0pt}{5mm} \\ &+2 \textbf{q}^{24} \textbf{t}^{14}+\textbf{q}^{26} \textbf{t}^{14}+\textbf{q}^{28} \textbf{t}^{14}+\textbf{q}^{28} \textbf{t}^{16}+\textbf{q}^{30} \textbf{t}^{16}+\textbf{q}^{32} \textbf{t}^{16}+\textbf{q}^{36} \textbf{t}^{18} & \rule{0pt}{5mm}  \\
\hline \textbf{a}^2 & \textbf{q}^2 \textbf{t}^3+\textbf{q}^4 \textbf{t}^5+\textbf{q}^6 \textbf{t}^5+\textbf{q}^6 \textbf{t}^7+2 \textbf{q}^8 \textbf{t}^7+\textbf{q}^{10} \textbf{t}^7+3 \textbf{q}^{10} \textbf{t}^9+2 \textbf{q}^{12} \textbf{t}^9+2 \textbf{q}^{12} \textbf{t}^{11}+\textbf{q}^{14} \textbf{t}^9+ & \rule{0pt}{5mm} \\ &+4 \textbf{q}^{14} \textbf{t}^{11}+2 \textbf{q}^{16} \textbf{t}^{11}+3 \textbf{q}^{16} \textbf{t}^{13}+\textbf{q}^{18} \textbf{t}^{11}+4 \textbf{q}^{18} \textbf{t}^{13}+\textbf{q}^{18} \textbf{t}^{15}+2 \textbf{q}^{20} \textbf{t}^{13}+3 \textbf{q}^{20} \textbf{t}^{15}+ & \rule{0pt}{5mm} \\ &+\textbf{q}^{22} \textbf{t}^{13}+4 \textbf{q}^{22} \textbf{t}^{15}+2 \textbf{q}^{24} \textbf{t}^{15}+2 \textbf{q}^{24} \textbf{t}^{17}+\textbf{q}^{26} \textbf{t}^{15}+3 \textbf{q}^{26} \textbf{t}^{17}+2 \textbf{q}^{28} \textbf{t}^{17}+\textbf{q}^{30} \textbf{t}^{17}+ & \rule{0pt}{5mm} \\ &+\textbf{q}^{30} \textbf{t}^{19}+\textbf{q}^{32} \textbf{t}^{19}+\textbf{q}^{34} \textbf{t}^{19} & \rule{0pt}{5mm}  \\
\hline \textbf{a}^4 & \textbf{q}^6 \textbf{t}^8+\textbf{q}^8 \textbf{t}^{10}+\textbf{q}^{10} \textbf{t}^{10}+\textbf{q}^{10} \textbf{t}^{12}+2 \textbf{q}^{12} \textbf{t}^{12}+\textbf{q}^{14} \textbf{t}^{12}+3 \textbf{q}^{14} \textbf{t}^{14}+2 \textbf{q}^{16} \textbf{t}^{14}+\textbf{q}^{16} \textbf{t}^{16}+ & \rule{0pt}{5mm} \\ &+\textbf{q}^{18} \textbf{t}^{14}+3 \textbf{q}^{18} \textbf{t}^{16}+2 \textbf{q}^{20} \textbf{t}^{16}+\textbf{q}^{20} \textbf{t}^{18}+\textbf{q}^{22} \textbf{t}^{16}+3 \textbf{q}^{22} \textbf{t}^{18}+2 \textbf{q}^{24} \textbf{t}^{18}+\textbf{q}^{26} \textbf{t}^{18}+ & \rule{0pt}{5mm} \\ &+\textbf{q}^{26} \textbf{t}^{20}+\textbf{q}^{28} \textbf{t}^{20}+\textbf{q}^{30} \textbf{t}^{20} & \rule{0pt}{5mm}  \\
\hline \textbf{a}^6 & \textbf{q}^{12} \textbf{t}^{15}+\textbf{q}^{16} \textbf{t}^{17}+\textbf{q}^{18} \textbf{t}^{19}+\textbf{q}^{20} \textbf{t}^{19}+\textbf{q}^{24} \textbf{t}^{21} & \rule{0pt}{5mm}  \\
\end{array}
\]

\paragraph{The case $(n,m) = (4,11)$}

\[
\begin{array}{c|lll}
\textbf{a}-{\rm degree} & {\rm coefficient} & \rule{0pt}{3mm}  \\
\hline \textbf{a}^0 & 1+\textbf{q}^4 \textbf{t}^2+\textbf{q}^6 \textbf{t}^4+\textbf{q}^8 \textbf{t}^4+\textbf{q}^8 \textbf{t}^6+\textbf{q}^{10} \textbf{t}^6+\textbf{q}^{12} \textbf{t}^6+2 \textbf{q}^{12} \textbf{t}^8+\textbf{q}^{14} \textbf{t}^8+\textbf{q}^{14} \textbf{t}^{10}+ & \rule{0pt}{5mm} \\ &+\textbf{q}^{16} \textbf{t}^8+2 \textbf{q}^{16} \textbf{t}^{10}+\textbf{q}^{16} \textbf{t}^{12}+\textbf{q}^{18} \textbf{t}^{10}+2 \textbf{q}^{18} \textbf{t}^{12}+\textbf{q}^{20} \textbf{t}^{10}+2 \textbf{q}^{20} \textbf{t}^{12}+2 \textbf{q}^{20} \textbf{t}^{14}+ & \rule{0pt}{5mm} \\ &+\textbf{q}^{22} \textbf{t}^{12}+2 \textbf{q}^{22} \textbf{t}^{14}+\textbf{q}^{24} \textbf{t}^{12}+\textbf{q}^{22} \textbf{t}^{16}+2 \textbf{q}^{24} \textbf{t}^{14}+3 \textbf{q}^{24} \textbf{t}^{16}+\textbf{q}^{26} \textbf{t}^{14}+2 \textbf{q}^{26} \textbf{t}^{16}+ & \rule{0pt}{5mm} \\ &+\textbf{q}^{28} \textbf{t}^{14}+\textbf{q}^{26} \textbf{t}^{18}+2 \textbf{q}^{28} \textbf{t}^{16}+3 \textbf{q}^{28} \textbf{t}^{18}+\textbf{q}^{30} \textbf{t}^{16}+2 \textbf{q}^{30} \textbf{t}^{18}+\textbf{q}^{32} \textbf{t}^{16}+2 \textbf{q}^{30} \textbf{t}^{20}+ & \rule{0pt}{5mm} \\ &+2 \textbf{q}^{32} \textbf{t}^{18}+3 \textbf{q}^{32} \textbf{t}^{20}+\textbf{q}^{34} \textbf{t}^{18}+2 \textbf{q}^{34} \textbf{t}^{20}+\textbf{q}^{36} \textbf{t}^{18}+\textbf{q}^{34} \textbf{t}^{22}+2 \textbf{q}^{36} \textbf{t}^{20}+3 \textbf{q}^{36} \textbf{t}^{22}+ & \rule{0pt}{5mm} \\ &+\textbf{q}^{38} \textbf{t}^{20}+2 \textbf{q}^{38} \textbf{t}^{22}+\textbf{q}^{40} \textbf{t}^{20}+\textbf{q}^{38} \textbf{t}^{24}+2 \textbf{q}^{40} \textbf{t}^{22}+2 \textbf{q}^{40} \textbf{t}^{24}+\textbf{q}^{42} \textbf{t}^{22}+2 \textbf{q}^{42} \textbf{t}^{24}+ & \rule{0pt}{5mm} \\ &+\textbf{q}^{44} \textbf{t}^{22}+2 \textbf{q}^{44} \textbf{t}^{24}+\textbf{q}^{44} \textbf{t}^{26}+\textbf{q}^{46} \textbf{t}^{24}+\textbf{q}^{46} \textbf{t}^{26}+\textbf{q}^{48} \textbf{t}^{24}+2 \textbf{q}^{48} \textbf{t}^{26}+\textbf{q}^{50} \textbf{t}^{26}+ & \rule{0pt}{5mm} \\ &+\textbf{q}^{52} \textbf{t}^{26}+\textbf{q}^{52} \textbf{t}^{28}+\textbf{q}^{54} \textbf{t}^{28}+\textbf{q}^{56} \textbf{t}^{28}+\textbf{q}^{60} \textbf{t}^{30} & \rule{0pt}{5mm}  \\
\hline \textbf{a}^2 & \textbf{q}^2 \textbf{t}^3+\textbf{q}^4 \textbf{t}^5+\textbf{q}^6 \textbf{t}^5+\textbf{q}^6 \textbf{t}^7+2 \textbf{q}^8 \textbf{t}^7+\textbf{q}^{10} \textbf{t}^7+3 \textbf{q}^{10} \textbf{t}^9+2 \textbf{q}^{12} \textbf{t}^9+2 \textbf{q}^{12} \textbf{t}^{11}+\textbf{q}^{14} \textbf{t}^9+ & \rule{0pt}{5mm} \\ &+4 \textbf{q}^{14} \textbf{t}^{11}+\textbf{q}^{14} \textbf{t}^{13}+2 \textbf{q}^{16} \textbf{t}^{11}+4 \textbf{q}^{16} \textbf{t}^{13}+\textbf{q}^{18} \textbf{t}^{11}+4 \textbf{q}^{18} \textbf{t}^{13}+4 \textbf{q}^{18} \textbf{t}^{15}+2 \textbf{q}^{20} \textbf{t}^{13}+ & \rule{0pt}{5mm} \\ &+5 \textbf{q}^{20} \textbf{t}^{15}+\textbf{q}^{22} \textbf{t}^{13}+2 \textbf{q}^{20} \textbf{t}^{17}+4 \textbf{q}^{22} \textbf{t}^{15}+6 \textbf{q}^{22} \textbf{t}^{17}+2 \textbf{q}^{24} \textbf{t}^{15}+5 \textbf{q}^{24} \textbf{t}^{17}+\textbf{q}^{26} \textbf{t}^{15}+ & \rule{0pt}{5mm} \\ &+4 \textbf{q}^{24} \textbf{t}^{19}+4 \textbf{q}^{26} \textbf{t}^{17}+7 \textbf{q}^{26} \textbf{t}^{19}+2 \textbf{q}^{28} \textbf{t}^{17}+\textbf{q}^{26} \textbf{t}^{21}+5 \textbf{q}^{28} \textbf{t}^{19}+\textbf{q}^{30} \textbf{t}^{17}+5 \textbf{q}^{28} \textbf{t}^{21}+ & \rule{0pt}{5mm} \\ &+4 \textbf{q}^{30} \textbf{t}^{19}+7 \textbf{q}^{30} \textbf{t}^{21}+2 \textbf{q}^{32} \textbf{t}^{19}+\textbf{q}^{30} \textbf{t}^{23}+5 \textbf{q}^{32} \textbf{t}^{21}+\textbf{q}^{34} \textbf{t}^{19}+5 \textbf{q}^{32} \textbf{t}^{23}+4 \textbf{q}^{34} \textbf{t}^{21}+ & \rule{0pt}{5mm} \\ &+7 \textbf{q}^{34} \textbf{t}^{23}+2 \textbf{q}^{36} \textbf{t}^{21}+\textbf{q}^{34} \textbf{t}^{25}+5 \textbf{q}^{36} \textbf{t}^{23}+\textbf{q}^{38} \textbf{t}^{21}+4 \textbf{q}^{36} \textbf{t}^{25}+4 \textbf{q}^{38} \textbf{t}^{23}+6 \textbf{q}^{38} \textbf{t}^{25}+ & \rule{0pt}{5mm} \\ &+2 \textbf{q}^{40} \textbf{t}^{23}+5 \textbf{q}^{40} \textbf{t}^{25}+\textbf{q}^{42} \textbf{t}^{23}+2 \textbf{q}^{40} \textbf{t}^{27}+4 \textbf{q}^{42} \textbf{t}^{25}+4 \textbf{q}^{42} \textbf{t}^{27}+2 \textbf{q}^{44} \textbf{t}^{25}+4 \textbf{q}^{44} \textbf{t}^{27}+ & \rule{0pt}{5mm} \\ &+\textbf{q}^{46} \textbf{t}^{25}+4 \textbf{q}^{46} \textbf{t}^{27}+\textbf{q}^{46} \textbf{t}^{29}+2 \textbf{q}^{48} \textbf{t}^{27}+2 \textbf{q}^{48} \textbf{t}^{29}+\textbf{q}^{50} \textbf{t}^{27}+3 \textbf{q}^{50} \textbf{t}^{29}+2 \textbf{q}^{52} \textbf{t}^{29}+ & \rule{0pt}{5mm} \\ &+\textbf{q}^{54} \textbf{t}^{29}+\textbf{q}^{54} \textbf{t}^{31}+\textbf{q}^{56} \textbf{t}^{31}+\textbf{q}^{58} \textbf{t}^{31} & \rule{0pt}{5mm}  \\
\hline \textbf{a}^4 & \textbf{q}^6 \textbf{t}^8+\textbf{q}^8 \textbf{t}^{10}+\textbf{q}^{10} \textbf{t}^{10}+\textbf{q}^{10} \textbf{t}^{12}+2 \textbf{q}^{12} \textbf{t}^{12}+\textbf{q}^{14} \textbf{t}^{12}+3 \textbf{q}^{14} \textbf{t}^{14}+2 \textbf{q}^{16} \textbf{t}^{14}+2 \textbf{q}^{16} \textbf{t}^{16}+ & \rule{0pt}{5mm} \\ &+\textbf{q}^{18} \textbf{t}^{14}+4 \textbf{q}^{18} \textbf{t}^{16}+\textbf{q}^{18} \textbf{t}^{18}+2 \textbf{q}^{20} \textbf{t}^{16}+4 \textbf{q}^{20} \textbf{t}^{18}+\textbf{q}^{22} \textbf{t}^{16}+4 \textbf{q}^{22} \textbf{t}^{18}+4 \textbf{q}^{22} \textbf{t}^{20}+ & \rule{0pt}{5mm} \\ &+2 \textbf{q}^{24} \textbf{t}^{18}+5 \textbf{q}^{24} \textbf{t}^{20}+\textbf{q}^{26} \textbf{t}^{18}+\textbf{q}^{24} \textbf{t}^{22}+4 \textbf{q}^{26} \textbf{t}^{20}+5 \textbf{q}^{26} \textbf{t}^{22}+2 \textbf{q}^{28} \textbf{t}^{20}+5 \textbf{q}^{28} \textbf{t}^{22}+ & \rule{0pt}{5mm} \\ &+\textbf{q}^{30} \textbf{t}^{20}+2 \textbf{q}^{28} \textbf{t}^{24}+4 \textbf{q}^{30} \textbf{t}^{22}+6 \textbf{q}^{30} \textbf{t}^{24}+2 \textbf{q}^{32} \textbf{t}^{22}+5 \textbf{q}^{32} \textbf{t}^{24}+\textbf{q}^{34} \textbf{t}^{22}+2 \textbf{q}^{32} \textbf{t}^{26}+ & \rule{0pt}{5mm} \\ &+4 \textbf{q}^{34} \textbf{t}^{24}+5 \textbf{q}^{34} \textbf{t}^{26}+2 \textbf{q}^{36} \textbf{t}^{24}+5 \textbf{q}^{36} \textbf{t}^{26}+\textbf{q}^{38} \textbf{t}^{24}+\textbf{q}^{36} \textbf{t}^{28}+4 \textbf{q}^{38} \textbf{t}^{26}+4 \textbf{q}^{38} \textbf{t}^{28}+ & \rule{0pt}{5mm} \\ &+2 \textbf{q}^{40} \textbf{t}^{26}+4 \textbf{q}^{40} \textbf{t}^{28}+\textbf{q}^{42} \textbf{t}^{26}+4 \textbf{q}^{42} \textbf{t}^{28}+\textbf{q}^{42} \textbf{t}^{30}+2 \textbf{q}^{44} \textbf{t}^{28}+2 \textbf{q}^{44} \textbf{t}^{30}+\textbf{q}^{46} \textbf{t}^{28}+ & \rule{0pt}{5mm} \\ &+3 \textbf{q}^{46} \textbf{t}^{30}+2 \textbf{q}^{48} \textbf{t}^{30}+\textbf{q}^{50} \textbf{t}^{30}+\textbf{q}^{50} \textbf{t}^{32}+\textbf{q}^{52} \textbf{t}^{32}+\textbf{q}^{54} \textbf{t}^{32} & \rule{0pt}{5mm}  \\
\hline \textbf{a}^6 & \textbf{q}^{12} \textbf{t}^{15}+\textbf{q}^{16} \textbf{t}^{17}+\textbf{q}^{18} \textbf{t}^{19}+\textbf{q}^{20} \textbf{t}^{19}+\textbf{q}^{20} \textbf{t}^{21}+\textbf{q}^{22} \textbf{t}^{21}+\textbf{q}^{24} \textbf{t}^{21}+2 \textbf{q}^{24} \textbf{t}^{23}+ & \rule{0pt}{5mm} \\ &+\textbf{q}^{26} \textbf{t}^{23}+\textbf{q}^{26} \textbf{t}^{25}+\textbf{q}^{28} \textbf{t}^{23}+2 \textbf{q}^{28} \textbf{t}^{25}+\textbf{q}^{30} \textbf{t}^{25}+\textbf{q}^{30} \textbf{t}^{27}+\textbf{q}^{32} \textbf{t}^{25}+ & \rule{0pt}{5mm} \\ &+2 \textbf{q}^{32} \textbf{t}^{27}+\textbf{q}^{34} \textbf{t}^{27}+\textbf{q}^{34} \textbf{t}^{29}+\textbf{q}^{36} \textbf{t}^{27}+2 \textbf{q}^{36} \textbf{t}^{29}+\textbf{q}^{38} \textbf{t}^{29}+\textbf{q}^{40} \textbf{t}^{29}+ & \rule{0pt}{5mm} \\ &+\textbf{q}^{40} \textbf{t}^{31}+\textbf{q}^{42} \textbf{t}^{31}+\textbf{q}^{44} \textbf{t}^{31}+\textbf{q}^{48} \textbf{t}^{33} & \rule{0pt}{5mm}  \\
\end{array}
\]

\subsection*{The case $(n,m) = (5,6)$}

\[
\begin{array}{c|lll}
\textbf{a}-{\rm degree} & {\rm coefficient} & \rule{0pt}{3mm}  \\
\hline \textbf{a}^0 & 1+\textbf{q}^4 \textbf{t}^2+\textbf{q}^6 \textbf{t}^4+\textbf{q}^8 \textbf{t}^4+\textbf{q}^8 \textbf{t}^6+\textbf{q}^{10} \textbf{t}^6+\textbf{q}^{10} \textbf{t}^8+\textbf{q}^{12} \textbf{t}^6+2 \textbf{q}^{12} \textbf{t}^8+\textbf{q}^{14} \textbf{t}^8+ & \rule{0pt}{5mm} \\ &+\textbf{q}^{14} \textbf{t}^{10}+\textbf{q}^{16} \textbf{t}^8+2 \textbf{q}^{16} \textbf{t}^{10}+\textbf{q}^{16} \textbf{t}^{12}+\textbf{q}^{18} \textbf{t}^{10}+2 \textbf{q}^{18} \textbf{t}^{12}+\textbf{q}^{20} \textbf{t}^{10}+2 \textbf{q}^{20} \textbf{t}^{12}+ & \rule{0pt}{5mm} \\ &+\textbf{q}^{20} \textbf{t}^{14}+\textbf{q}^{22} \textbf{t}^{12}+2 \textbf{q}^{22} \textbf{t}^{14}+\textbf{q}^{24} \textbf{t}^{12}+2 \textbf{q}^{24} \textbf{t}^{14}+\textbf{q}^{24} \textbf{t}^{16}+\textbf{q}^{26} \textbf{t}^{14}+\textbf{q}^{26} \textbf{t}^{16} + & \rule{0pt}{5mm} \\ &+ \textbf{q}^{28} \textbf{t}^{14}+2 \textbf{q}^{28} \textbf{t}^{16}+\textbf{q}^{30} \textbf{t}^{16}+\textbf{q}^{30} \textbf{t}^{18}+\textbf{q}^{32} \textbf{t}^{16}+\textbf{q}^{32} \textbf{t}^{18} + & \rule{0pt}{5mm} \\ & + \textbf{q}^{34} \textbf{t}^{18}+\textbf{q}^{36} \textbf{t}^{18}+\textbf{q}^{40} \textbf{t}^{20} & \rule{0pt}{5mm} \\
\hline \textbf{a}^2 & \textbf{q}^2 \textbf{t}^3+\textbf{q}^4 \textbf{t}^5+\textbf{q}^6 \textbf{t}^5+\textbf{q}^6 \textbf{t}^7+2 \textbf{q}^8 \textbf{t}^7+\textbf{q}^8 \textbf{t}^9+\textbf{q}^{10} \textbf{t}^7+3 \textbf{q}^{10} \textbf{t}^9+2 \textbf{q}^{12} \textbf{t}^9+3 \textbf{q}^{12} \textbf{t}^{11}+ & \rule{0pt}{5mm} \\ & + \textbf{q}^{14} \textbf{t}^9+4 \textbf{q}^{14} \textbf{t}^{11}+2 \textbf{q}^{14} \textbf{t}^{13}+2 \textbf{q}^{16} \textbf{t}^{11}+4 \textbf{q}^{16} \textbf{t}^{13}+\textbf{q}^{18} \textbf{t}^{11}+\textbf{q}^{16} \textbf{t}^{15}+4 \textbf{q}^{18} \textbf{t}^{13}+ & \rule{0pt}{5mm} \\ & + 3 \textbf{q}^{18} \textbf{t}^{15}+2 \textbf{q}^{20} \textbf{t}^{13}+5 \textbf{q}^{20} \textbf{t}^{15}+\textbf{q}^{22} \textbf{t}^{13}+\textbf{q}^{20} \textbf{t}^{17}+4 \textbf{q}^{22} \textbf{t}^{15}+3 \textbf{q}^{22} \textbf{t}^{17}+ & \rule{0pt}{5mm} \\ & + 2 \textbf{q}^{24} \textbf{t}^{15}+4 \textbf{q}^{24} \textbf{t}^{17}+\textbf{q}^{26} \textbf{t}^{15}+\textbf{q}^{24} \textbf{t}^{19}+4 \textbf{q}^{26} \textbf{t}^{17}+2 \textbf{q}^{26} \textbf{t}^{19}+2 \textbf{q}^{28} \textbf{t}^{17}+ & \rule{0pt}{5mm} \\ & + 3 \textbf{q}^{28} \textbf{t}^{19}+\textbf{q}^{30} \textbf{t}^{17}+3 \textbf{q}^{30} \textbf{t}^{19}+2 \textbf{q}^{32} \textbf{t}^{19}+\textbf{q}^{32} \textbf{t}^{21}+\textbf{q}^{34} \textbf{t}^{19}+ & \rule{0pt}{5mm} \\ & + \textbf{q}^{34} \textbf{t}^{21}+\textbf{q}^{36} \textbf{t}^{21}+\textbf{q}^{38} \textbf{t}^{21} \rule{0pt}{5mm} \\
\hline \textbf{a}^4 & \textbf{q}^6 \textbf{t}^8+\textbf{q}^8 \textbf{t}^{10}+\textbf{q}^{10} \textbf{t}^{10}+2 \textbf{q}^{10} \textbf{t}^{12}+2 \textbf{q}^{12} \textbf{t}^{12}+\textbf{q}^{12} \textbf{t}^{14}+\textbf{q}^{14} \textbf{t}^{12}+3 \textbf{q}^{14} \textbf{t}^{14}+ & \rule{0pt}{5mm} \\ & +\textbf{q}^{14} \textbf{t}^{16}+2 \textbf{q}^{16} \textbf{t}^{14}+3 \textbf{q}^{16} \textbf{t}^{16}+\textbf{q}^{18} \textbf{t}^{14}+4 \textbf{q}^{18} \textbf{t}^{16}+2 \textbf{q}^{18} \textbf{t}^{18}+2 \textbf{q}^{20} \textbf{t}^{16}+ & \rule{0pt}{5mm} \\ & +3 \textbf{q}^{20} \textbf{t}^{18}+\textbf{q}^{22} \textbf{t}^{16}+\textbf{q}^{20} \textbf{t}^{20}+4 \textbf{q}^{22} \textbf{t}^{18}+2 \textbf{q}^{22} \textbf{t}^{20}+2 \textbf{q}^{24} \textbf{t}^{18}+3 \textbf{q}^{24} \textbf{t}^{20}+ & \rule{0pt}{5mm} \\ & +\textbf{q}^{26} \textbf{t}^{18}+3 \textbf{q}^{26} \textbf{t}^{20}+\textbf{q}^{26} \textbf{t}^{22}+2 \textbf{q}^{28} \textbf{t}^{20}+\textbf{q}^{28} \textbf{t}^{22}+\textbf{q}^{30} \textbf{t}^{20}+ & \rule{0pt}{5mm} \\ & +2 \textbf{q}^{30} \textbf{t}^{22}+\textbf{q}^{32} \textbf{t}^{22}+\textbf{q}^{34} \textbf{t}^{22} \rule{0pt}{5mm} \\
\hline \textbf{a}^6 & \textbf{q}^{12} \textbf{t}^{15}+\textbf{q}^{14} \textbf{t}^{17}+\textbf{q}^{16} \textbf{t}^{17}+\textbf{q}^{16} \textbf{t}^{19}+\textbf{q}^{18} \textbf{t}^{19}+\textbf{q}^{18} \textbf{t}^{21}+\textbf{q}^{20} \textbf{t}^{19}+\textbf{q}^{20} \textbf{t}^{21}+ & \rule{0pt}{5mm} \\ & +\textbf{q}^{22} \textbf{t}^{21}+\textbf{q}^{22} \textbf{t}^{23}+\textbf{q}^{24} \textbf{t}^{21}+\textbf{q}^{24} \textbf{t}^{23}+\textbf{q}^{26} \textbf{t}^{23}+\textbf{q}^{28} \textbf{t}^{23} \rule{0pt}{5mm} \\
\hline \textbf{a}^8 & \textbf{t}^{24} \textbf{q}^{20} \rule{0pt}{5mm}
\end{array}
\]

\vspace{-1ex}

\subsection*{The case $(n,m) = (5,8)$}

\vspace{-1ex}

\[
\begin{array}{c|lll}
\textbf{a}-{\rm degree} & {\rm coefficient} & \rule{0pt}{3mm}  \\
\hline 
\textbf{a}^0 & 1+\textbf{q}^4 \textbf{t}^2+\textbf{q}^4 \textbf{t}^4+\textbf{q}^6 \textbf{t}^4+\textbf{q}^8 \textbf{t}^4+2 \textbf{q}^8 \textbf{t}^6+\textbf{q}^{10} \textbf{t}^6+2 \textbf{q}^{10} \textbf{t}^8+\textbf{q}^{12} \textbf{t}^6+3 \textbf{q}^{12} \textbf{t}^8+\textbf{q}^{12} \textbf{t}^{10}+ & \rule{0pt}{5mm} \\ & +\textbf{q}^{14} \textbf{t}^8+3 \textbf{q}^{14} \textbf{t}^{10}+\textbf{q}^{16} \textbf{t}^8+\textbf{q}^{14} \textbf{t}^{12}+3 \textbf{q}^{16} \textbf{t}^{10}+4 \textbf{q}^{16} \textbf{t}^{12}+\textbf{q}^{18} \textbf{t}^{10}+4 \textbf{q}^{18} \textbf{t}^{12}+\textbf{q}^{20} \textbf{t}^{10}+ & \rule{0pt}{5mm} \\ & +2 \textbf{q}^{18} \textbf{t}^{14}+3 \textbf{q}^{20} \textbf{t}^{12}+5 \textbf{q}^{20} \textbf{t}^{14}+\textbf{q}^{22} \textbf{t}^{12}+2 \textbf{q}^{20} \textbf{t}^{16}+4 \textbf{q}^{22} \textbf{t}^{14}+\textbf{q}^{24} \textbf{t}^{12}+4 \textbf{q}^{22} \textbf{t}^{16}+ & \rule{0pt}{5mm} \\ & +3 \textbf{q}^{24} \textbf{t}^{14}+6 \textbf{q}^{24} \textbf{t}^{16}+\textbf{q}^{26} \textbf{t}^{14}+3 \textbf{q}^{24} \textbf{t}^{18}+4 \textbf{q}^{26} \textbf{t}^{16}+\textbf{q}^{28} \textbf{t}^{14}+5 \textbf{q}^{26} \textbf{t}^{18}+3 \textbf{q}^{28} \textbf{t}^{16}+ & \rule{0pt}{5mm} \\ & +\textbf{q}^{26} \textbf{t}^{20}+6 \textbf{q}^{28} \textbf{t}^{18}+\textbf{q}^{30} \textbf{t}^{16}+4 \textbf{q}^{28} \textbf{t}^{20}+4 \textbf{q}^{30} \textbf{t}^{18}+\textbf{q}^{32} \textbf{t}^{16}+6 \textbf{q}^{30} \textbf{t}^{20}+3 \textbf{q}^{32} \textbf{t}^{18}+ & \rule{0pt}{5mm} \\ & +\textbf{q}^{30} \textbf{t}^{22}+6 \textbf{q}^{32} \textbf{t}^{20}+\textbf{q}^{34} \textbf{t}^{18}+4 \textbf{q}^{32} \textbf{t}^{22}+4 \textbf{q}^{34} \textbf{t}^{20}+\textbf{q}^{36} \textbf{t}^{18}+5 \textbf{q}^{34} \textbf{t}^{22}+3 \textbf{q}^{36} \textbf{t}^{20}+ & \rule{0pt}{5mm} \\ & +\textbf{q}^{34} \textbf{t}^{24}+6 \textbf{q}^{36} \textbf{t}^{22}+\textbf{q}^{38} \textbf{t}^{20}+3 \textbf{q}^{36} \textbf{t}^{24}+4 \textbf{q}^{38} \textbf{t}^{22}+\textbf{q}^{40} \textbf{t}^{20}+4 \textbf{q}^{38} \textbf{t}^{24}+3 \textbf{q}^{40} \textbf{t}^{22}+ & \rule{0pt}{5mm} \\ & +5 \textbf{q}^{40} \textbf{t}^{24}+\textbf{q}^{42} \textbf{t}^{22}+2 \textbf{q}^{40} \textbf{t}^{26}+4 \textbf{q}^{42} \textbf{t}^{24}+\textbf{q}^{44} \textbf{t}^{22}+2 \textbf{q}^{42} \textbf{t}^{26}+3 \textbf{q}^{44} \textbf{t}^{24}+4 \textbf{q}^{44} \textbf{t}^{26}+ & \rule{0pt}{5mm} \\ & +\textbf{q}^{46} \textbf{t}^{24}+3 \textbf{q}^{46} \textbf{t}^{26}+\textbf{q}^{48} \textbf{t}^{24}+\textbf{q}^{46} \textbf{t}^{28}+3 \textbf{q}^{48} \textbf{t}^{26}+\textbf{q}^{48} \textbf{t}^{28}+\textbf{q}^{50} \textbf{t}^{26}+2 \textbf{q}^{50} \textbf{t}^{28}+ & \rule{0pt}{5mm} \\ & +\textbf{q}^{52} \textbf{t}^{26}+2 \textbf{q}^{52} \textbf{t}^{28}+\textbf{q}^{54} \textbf{t}^{28}+\textbf{q}^{56} \textbf{t}^{28}+\textbf{q}^{56} \textbf{t}^{30}+\textbf{q}^{60} \textbf{t}^{30} \rule{0pt}{5mm} \\
\end{array}
\]
\emph{}
\vspace{-20ex}
\[
\begin{array}{c|lll}
\textbf{a}-{\rm degree} & {\rm coefficient} & \rule{0pt}{3mm}  \\
\hline \textbf{a}^2 & \textbf{q}^2 \textbf{t}^3+\textbf{q}^4 \textbf{t}^5+\textbf{q}^6 \textbf{t}^5+2 \textbf{q}^6 \textbf{t}^7+2 \textbf{q}^8 \textbf{t}^7+2 \textbf{q}^8 \textbf{t}^9+\textbf{q}^{10} \textbf{t}^7+4 \textbf{q}^{10} \textbf{t}^9+\textbf{q}^{10} \textbf{t}^{11}+2 \textbf{q}^{12} \textbf{t}^9+ & \rule{0pt}{5mm} \\ & +6 \textbf{q}^{12} \textbf{t}^{11}+\textbf{q}^{14} \textbf{t}^9+\textbf{q}^{12} \textbf{t}^{13}+5 \textbf{q}^{14} \textbf{t}^{11}+6 \textbf{q}^{14} \textbf{t}^{13}+2 \textbf{q}^{16} \textbf{t}^{11}+8 \textbf{q}^{16} \textbf{t}^{13}+\textbf{q}^{18} \textbf{t}^{11}+ & \rule{0pt}{5mm} \\ & +5 \textbf{q}^{16} \textbf{t}^{15}+5 \textbf{q}^{18} \textbf{t}^{13}+11 \textbf{q}^{18} \textbf{t}^{15}+2 \textbf{q}^{20} \textbf{t}^{13}+3 \textbf{q}^{18} \textbf{t}^{17}+9 \textbf{q}^{20} \textbf{t}^{15}+\textbf{q}^{22} \textbf{t}^{13}+ & \rule{0pt}{5mm} \\ & +10 \textbf{q}^{20} \textbf{t}^{17}+5 \textbf{q}^{22} \textbf{t}^{15}+\textbf{q}^{20} \textbf{t}^{19}+13 \textbf{q}^{22} \textbf{t}^{17}+2 \textbf{q}^{24} \textbf{t}^{15}+8 \textbf{q}^{22} \textbf{t}^{19}+9 \textbf{q}^{24} \textbf{t}^{17}+ & \rule{0pt}{5mm} \\ & +\textbf{q}^{26} \textbf{t}^{15}+14 \textbf{q}^{24} \textbf{t}^{19}+5 \textbf{q}^{26} \textbf{t}^{17}+4 \textbf{q}^{24} \textbf{t}^{21}+14 \textbf{q}^{26} \textbf{t}^{19}+2 \textbf{q}^{28} \textbf{t}^{17}+11 \textbf{q}^{26} \textbf{t}^{21}+ & \rule{0pt}{5mm} \\ & +9 \textbf{q}^{28} \textbf{t}^{19}+\textbf{q}^{30} \textbf{t}^{17}+\textbf{q}^{26} \textbf{t}^{23}+16 \textbf{q}^{28} \textbf{t}^{21}+5 \textbf{q}^{30} \textbf{t}^{19}+6 \textbf{q}^{28} \textbf{t}^{23}+14 \textbf{q}^{30} \textbf{t}^{21}+ & \rule{0pt}{5mm} \\ & +2 \textbf{q}^{32} \textbf{t}^{19}+13 \textbf{q}^{30} \textbf{t}^{23}+9 \textbf{q}^{32} \textbf{t}^{21}+\textbf{q}^{34} \textbf{t}^{19}+\textbf{q}^{30} \textbf{t}^{25}+16 \textbf{q}^{32} \textbf{t}^{23}+5 \textbf{q}^{34} \textbf{t}^{21}+ & \rule{0pt}{5mm} \\ & +6 \textbf{q}^{32} \textbf{t}^{25}+14 \textbf{q}^{34} \textbf{t}^{23}+2 \textbf{q}^{36} \textbf{t}^{21}+11 \textbf{q}^{34} \textbf{t}^{25}+9 \textbf{q}^{36} \textbf{t}^{23}+\textbf{q}^{38} \textbf{t}^{21}+\textbf{q}^{34} \textbf{t}^{27}+ & \rule{0pt}{5mm} \\ & +14 \textbf{q}^{36} \textbf{t}^{25}+5 \textbf{q}^{38} \textbf{t}^{23}+4 \textbf{q}^{36} \textbf{t}^{27}+13 \textbf{q}^{38} \textbf{t}^{25}+2 \textbf{q}^{40} \textbf{t}^{23}+8 \textbf{q}^{38} \textbf{t}^{27}+9 \textbf{q}^{40} \textbf{t}^{25}+ & \rule{0pt}{5mm} \\ & +\textbf{q}^{42} \textbf{t}^{23}+10 \textbf{q}^{40} \textbf{t}^{27}+5 \textbf{q}^{42} \textbf{t}^{25}+\textbf{q}^{40} \textbf{t}^{29}+11 \textbf{q}^{42} \textbf{t}^{27}+2 \textbf{q}^{44} \textbf{t}^{25}+3 \textbf{q}^{42} \textbf{t}^{29}+ & \rule{0pt}{5mm} \\ & +8 \textbf{q}^{44} \textbf{t}^{27}+\textbf{q}^{46} \textbf{t}^{25}+5 \textbf{q}^{44} \textbf{t}^{29}+5 \textbf{q}^{46} \textbf{t}^{27}+6 \textbf{q}^{46} \textbf{t}^{29}+2 \textbf{q}^{48} \textbf{t}^{27}+6 \textbf{q}^{48} \textbf{t}^{29}+ & \rule{0pt}{5mm} \\ & +\textbf{q}^{50} \textbf{t}^{27}+\textbf{q}^{48} \textbf{t}^{31}+4 \textbf{q}^{50} \textbf{t}^{29}+\textbf{q}^{50} \textbf{t}^{31}+2 \textbf{q}^{52} \textbf{t}^{29}+2 \textbf{q}^{52} \textbf{t}^{31}+\textbf{q}^{54} \textbf{t}^{29}+ & \rule{0pt}{5mm} \\ & + 2 \textbf{q}^{54} \textbf{t}^{31}+\textbf{q}^{56} \textbf{t}^{31}+\textbf{q}^{58} \textbf{t}^{31} \rule{0pt}{5mm} \\
\hline \textbf{a}^4 & \textbf{q}^6 \textbf{t}^8+\textbf{q}^8 \textbf{t}^{10}+\textbf{q}^{10} \textbf{t}^{10}+3 \textbf{q}^{10} \textbf{t}^{12}+2 \textbf{q}^{12} \textbf{t}^{12}+2 \textbf{q}^{12} \textbf{t}^{14}+\textbf{q}^{14} \textbf{t}^{12}+5 \textbf{q}^{14} \textbf{t}^{14}+ & \rule{0pt}{5mm} \\ & +3 \textbf{q}^{14} \textbf{t}^{16}+2 \textbf{q}^{16} \textbf{t}^{14}+7 \textbf{q}^{16} \textbf{t}^{16}+\textbf{q}^{18} \textbf{t}^{14}+\textbf{q}^{16} \textbf{t}^{18}+6 \textbf{q}^{18} \textbf{t}^{16}+8 \textbf{q}^{18} \textbf{t}^{18}+ & \rule{0pt}{5mm} \\ & +2 \textbf{q}^{20} \textbf{t}^{16}+\textbf{q}^{18} \textbf{t}^{20}+9 \textbf{q}^{20} \textbf{t}^{18}+\textbf{q}^{22} \textbf{t}^{16}+7 \textbf{q}^{20} \textbf{t}^{20}+6 \textbf{q}^{22} \textbf{t}^{18}+13 \textbf{q}^{22} \textbf{t}^{20}+ & \rule{0pt}{5mm} \\ & +2 \textbf{q}^{24} \textbf{t}^{18}+4 \textbf{q}^{22} \textbf{t}^{22}+10 \textbf{q}^{24} \textbf{t}^{20}+\textbf{q}^{26} \textbf{t}^{18}+11 \textbf{q}^{24} \textbf{t}^{22}+6 \textbf{q}^{26} \textbf{t}^{20}+2 \textbf{q}^{24} \textbf{t}^{24}+ & \rule{0pt}{5mm} \\ & +15 \textbf{q}^{26} \textbf{t}^{22}+2 \textbf{q}^{28} \textbf{t}^{20}+9 \textbf{q}^{26} \textbf{t}^{24}+10 \textbf{q}^{28} \textbf{t}^{22}+\textbf{q}^{30} \textbf{t}^{20}+14 \textbf{q}^{28} \textbf{t}^{24}+6 \textbf{q}^{30} \textbf{t}^{22}+ & \rule{0pt}{5mm} \\ & +3 \textbf{q}^{28} \textbf{t}^{26}+16 \textbf{q}^{30} \textbf{t}^{24}+2 \textbf{q}^{32} \textbf{t}^{22}+10 \textbf{q}^{30} \textbf{t}^{26}+10 \textbf{q}^{32} \textbf{t}^{24}+\textbf{q}^{34} \textbf{t}^{22}+\textbf{q}^{30} \textbf{t}^{28}+ & \rule{0pt}{5mm} \\ & +14 \textbf{q}^{32} \textbf{t}^{26}+6 \textbf{q}^{34} \textbf{t}^{24}+3 \textbf{q}^{32} \textbf{t}^{28}+15 \textbf{q}^{34} \textbf{t}^{26}+2 \textbf{q}^{36} \textbf{t}^{24}+9 \textbf{q}^{34} \textbf{t}^{28}+10 \textbf{q}^{36} \textbf{t}^{26}+ & \rule{0pt}{5mm} \\ & +\textbf{q}^{38} \textbf{t}^{24}+11 \textbf{q}^{36} \textbf{t}^{28}+6 \textbf{q}^{38} \textbf{t}^{26}+2 \textbf{q}^{36} \textbf{t}^{30}+13 \textbf{q}^{38} \textbf{t}^{28}+2 \textbf{q}^{40} \textbf{t}^{26}+4 \textbf{q}^{38} \textbf{t}^{30}+ & \rule{0pt}{5mm} \\ & +9 \textbf{q}^{40} \textbf{t}^{28}+\textbf{q}^{42} \textbf{t}^{26}+7 \textbf{q}^{40} \textbf{t}^{30}+6 \textbf{q}^{42} \textbf{t}^{28}+8 \textbf{q}^{42} \textbf{t}^{30}+2 \textbf{q}^{44} \textbf{t}^{28}+\textbf{q}^{42} \textbf{t}^{32}+ & \rule{0pt}{5mm} \\ & +7 \textbf{q}^{44} \textbf{t}^{30}+\textbf{q}^{46} \textbf{t}^{28}+\textbf{q}^{44} \textbf{t}^{32}+5 \textbf{q}^{46} \textbf{t}^{30}+3 \textbf{q}^{46} \textbf{t}^{32}+2 \textbf{q}^{48} \textbf{t}^{30}+2 \textbf{q}^{48} \textbf{t}^{32}+ & \rule{0pt}{5mm} \\ & + \textbf{q}^{50} \textbf{t}^{30}+3 \textbf{q}^{50} \textbf{t}^{32}+\textbf{q}^{52} \textbf{t}^{32}+\textbf{q}^{54} \textbf{t}^{32} \rule{0pt}{5mm} \\
\hline \textbf{a}^6 & \textbf{q}^{12} \textbf{t}^{15}+\textbf{q}^{14} \textbf{t}^{17}+\textbf{q}^{16} \textbf{t}^{17}+2 \textbf{q}^{16} \textbf{t}^{19}+2 \textbf{q}^{18} \textbf{t}^{19}+2 \textbf{q}^{18} \textbf{t}^{21}+\textbf{q}^{20} \textbf{t}^{19}+4 \textbf{q}^{20} \textbf{t}^{21}+ & \rule{0pt}{5mm} \\ & +\textbf{q}^{20} \textbf{t}^{23}+2 \textbf{q}^{22} \textbf{t}^{21}+5 \textbf{q}^{22} \textbf{t}^{23}+\textbf{q}^{24} \textbf{t}^{21}+\textbf{q}^{22} \textbf{t}^{25}+5 \textbf{q}^{24} \textbf{t}^{23}+4 \textbf{q}^{24} \textbf{t}^{25}+2 \textbf{q}^{26} \textbf{t}^{23}+ & \rule{0pt}{5mm} \\ & +6 \textbf{q}^{26} \textbf{t}^{25}+\textbf{q}^{28} \textbf{t}^{23}+3 \textbf{q}^{26} \textbf{t}^{27}+5 \textbf{q}^{28} \textbf{t}^{25}+6 \textbf{q}^{28} \textbf{t}^{27}+2 \textbf{q}^{30} \textbf{t}^{25}+\textbf{q}^{28} \textbf{t}^{29}+7 \textbf{q}^{30} \textbf{t}^{27}+ & \rule{0pt}{5mm} \\ & +\textbf{q}^{32} \textbf{t}^{25}+3 \textbf{q}^{30} \textbf{t}^{29}+5 \textbf{q}^{32} \textbf{t}^{27}+6 \textbf{q}^{32} \textbf{t}^{29}+2 \textbf{q}^{34} \textbf{t}^{27}+\textbf{q}^{32} \textbf{t}^{31}+6 \textbf{q}^{34} \textbf{t}^{29}+\textbf{q}^{36} \textbf{t}^{27}+ & \rule{0pt}{5mm} \\ & +3 \textbf{q}^{34} \textbf{t}^{31}+5 \textbf{q}^{36} \textbf{t}^{29}+4 \textbf{q}^{36} \textbf{t}^{31}+2 \textbf{q}^{38} \textbf{t}^{29}+5 \textbf{q}^{38} \textbf{t}^{31}+\textbf{q}^{40} \textbf{t}^{29}+\textbf{q}^{38} \textbf{t}^{33}+4 \textbf{q}^{40} \textbf{t}^{31}+ & \rule{0pt}{5mm} \\ & +\textbf{q}^{40} \textbf{t}^{33}+2 \textbf{q}^{42} \textbf{t}^{31}+2 \textbf{q}^{42} \textbf{t}^{33}+\textbf{q}^{44} \textbf{t}^{31}+2 \textbf{q}^{44} \textbf{t}^{33}+\textbf{q}^{46} \textbf{t}^{33}+\textbf{q}^{48} \textbf{t}^{33} \rule{0pt}{5mm} \\
\hline \textbf{a}^8 & \textbf{q}^{20} \textbf{t}^{24}+\textbf{q}^{24} \textbf{t}^{26}+\textbf{q}^{24} \textbf{t}^{28}+\textbf{q}^{26} \textbf{t}^{28}+\textbf{q}^{28} \textbf{t}^{28}+\textbf{q}^{28} \textbf{t}^{30}+\textbf{q}^{30} \textbf{t}^{30}+\textbf{q}^{30} \textbf{t}^{32}+ & \rule{0pt}{5mm} \\ & +\textbf{q}^{32} \textbf{t}^{30}+\textbf{q}^{32} \textbf{t}^{32}+\textbf{q}^{34} \textbf{t}^{32}+\textbf{q}^{36} \textbf{t}^{32}+\textbf{q}^{36} \textbf{t}^{34}+\textbf{q}^{40} \textbf{t}^{34} \rule{0pt}{5mm}
\end{array}
\]
This is the first case not to fall into the $(n,nk+1)$ or $(n,nk+n-1)$ families; in this case, eq.(\ref{NewFormula}) does not agree with a superpolynomial of the (5,8) knot. \footnote{We thank E.Gorsky and A.Sleptsov for calculating the superpolynomial of the (5,8) knot.} Note that, unlike the correct (5,8) superpolynomial, this one also disagrees with refined Chern-Simons theory.

\section*{Acknowledgements}

We are grateful to Mina Aganagic, Eugene Gorsky, Sergei Gukov, Alexei Sleptsov and Andrei Smirnov for valuable discussions. This research is supported in part by Berkeley Center for Theoretical Physics, by the National Science Foundation (award number 0855653), by the Institute for the Physics and Mathematics of the Universe, by the US Department of Energy under Contract DE-AC02-05CH11231, by Ministry of Education and Science of the Russian Federation under contract 14.740.11.5194, by RFBR grant 10-01-00536 and by joint grants 09-02-93105-CNRSL, 09-02-91005-ANF.

\end{document}